 \documentclass[final,authoryear,5p,times,twocolumn]{elsarticle}

\usepackage{graphicx}
\usepackage{amssymb}
\usepackage{color}
\usepackage{url}
\usepackage{multirow}

\begin{document}

\begin{frontmatter}

\title{Testing the weak-form efficiency of the WTI crude oil futures market}
\author[BS,RCE]{Zhi-Qiang Jiang}
\author[BS,RCE,SS]{Wen-Jie Xie}
\author[BS,RCE,SS,ICCT]{Wei-Xing Zhou \corref{cor1}}
\cortext[cor1]{Corresponding author. Address: 130 Meilong Road, P.O. Box 114, School of Business,
              East China University of Science and Technology, Shanghai 200237, China, Phone: +86-21-64253634.}
\ead{wxzhou@ecust.edu.cn} %
\ead[url]{http://rce.ecust.edu.cn/}%

\address[BS]{School of Business, East China University of Science and Technology, Shanghai 200237, China}
\address[RCE]{Research Center for Econophysics, East China University of Science and Technology, Shanghai 200237, China}
\address[SS]{Department of Mathematics, East China University of Science and Technology, Shanghai 200237, China}
\address[ICCT]{Key Laboratory of Coal Gasification and Energy Chemical Engineering (MOE), East China University of Science and Technology, Shanghai 200237, China}

\begin{abstract}
We perform detrending moving average analysis (DMA) and detrended fluctuation analysis (DFA) of the WTI crude oil futures prices (1983-2012) to investigate its efficiency. We further put forward a strict statistical test in the spirit of bootstrapping to verify the weak-form market efficiency hypothesis by employing the DMA (or DFA) exponent as the statistic. We verify the weak-form efficiency of the crude oil futures market when the whole period is considered. When we break the whole series into three sub-series separated by the outbreaks of the Gulf War and the Iraq War, our statistical tests uncover that only the Gulf War has the impact of reducing the efficiency of the crude oil market. If we split the whole time series into two sub-series based on the signing date of the North American Free Trade Agreement, we find that the market is inefficient in the sub-periods during which the Gulf War broke out. We also perform the same analysis on short time series in moving windows and find that the market is inefficient only when some turbulent events occur, such as the oil price crash in 1985, the Gulf war, and the oil price crash in 2008. Our analysis may offer a new understanding of the efficiency of the crude oil futures market and shed new lights on the investigation of the efficiency in other financial markets.
\end{abstract}


\end{frontmatter}


\section{Introduction}
\label{S1:Introduction}

The unfolding financial crisis triggered by the U.S. subprime mortgage crisis in 2007 has been viewed as the most severe in the past centenary after the U.S. stock market crash in 1929. The latest financial crisis is followed by the global economic recession and the European sovereign debt crisis. During this period, we have witnessed the boom and bust of many financial and commodities bubbles \citep{Sornette-2009-IJTSE,Sornette-Woodard-2010}. The 2006-2008 crude oil bubble was a significant example \citep{Sornette-Woordard-Zhou-2009-PA}. The West Texas Intermediate (WTI) oil prices experienced an accelerated rise followed by a spectacular crash in July 2008. When measured in terms of daily close prices, the peak of USD 145.29 per barrel was reached on July 3, 2008 and a significant low of USD 33.87 was seen on December 19, 2008, which is a level not seen since 2004. The historical high was USD 147.2 per barrel, recorded in the middle of the day of July 11, 2008.

In May 2008 before the July peak, \cite{Sornette-Woordard-Zhou-2009-PA} utilized the log-periodic power-law (LPPL) models for financial bubbles \citep{Sornette-2003-PR,Sornette-2003} to analyze the oil prices and confirmed the presence of a speculative bubble. Furthermore, they successfully predicted the crash of the oil bubble with high precision. The original prediction was posted on the arXiv (0806.1170v1) on 6 June 2008. A few weeks later, \cite{Drozdz-Kwapien-Oswiecimka-2008-APPA} successfully predicted the Brent oil price crash (see arXiv:0802.4043v2, 24 June 2008) based on the LPPL models. \cite{Yu-Wang-Lai-2008-EE} proposed a neural network ensemble learning paradigm based on empirical mode decomposition to forecast crude oil prices and successfully predicted the turning point and the oil price plummet. More generally, there are a lot of studies showing that the oil prices are predictable both in turbulent periods and in normal phases. For instance, \cite{Shambora-Rossiter-2007-EE} used an artificial neural network model with moving average crossover inputs to predict prices in the crude oil futures market and uncovered significant profitability, and \cite{Wang-Yang-2010-EE} found significant out-of-sample predictability of intraday high-frequency (30-min) WTI crude oil futures prices in a bear phase (December 1, 2000 to November 31, 2001) and a bull phase (November 1, 2006 to October 31, 2007).

A direct consequence of the predictability of oil prices is that the oil futures market is inefficient in the weak form. In addition, the formation of bubbles and its bust imply that the degree of market inefficiency may evolve in a dynamic way. Indeed, quite a few studies have been conducted to directly test the efficiency of the oil markets, most of which devoted to estimate the Hurst index $H$ using different techniques based on the fact that a process with the Hurst index significantly differing 0.5 does not execute a random walk \citep{Mandelbrot-Wallis-1969c-WRR,Mandelbrot-1983,Mandelbrot-1997}. These investigations of oil market efficiency can be classified into two related categories. Some researchers studied oil prices in relatively long periods, while others put emphasis on the evolving behavior of market efficiency in moving windows of relatively short lengths{\footnote{It is interesting to note that \cite{Green-Mork-1991-JAEm} investigated the spot and ``official'' OPEC prices of crude oil for the sample period 1978-1985 and found that efficiency was rejected for the sample as a whole but with an improvement over time.}}. In the rest of this section, we will review the literature along these two lines and provide some discussions on existing works.

\subsection{Crude oil market efficiency over long period}

\cite{AlvarezRamirez-Cisneros-IbarraValdez-Soriano-2002-PA} performed R/S analysis on WTI and Brent crude oil prices over the period from November 1981 to March 2002. They found that $H=0.567$ for WTI crude oil prices and $H=0.585$ for Brent crude oil prices. Using the height-height correlation method (also called structure function approach in turbulence), they found that $H\approx0.47$ within the scaling range $[1,18]$.

\cite{Serletis-Andreadis-2004-EE} performed R/S analysis on the WTI crude oil prices over the period from 2 January 1990 to 28 February 2001 and found that $H=0.85$, which is a surprisingly large number indicating very strong autocorrelations (predictability) in the price time series. When they applied the spectral analysis on the prices in the frequency domain, the scaling exponent was found to be $\alpha=2.03$, which corresponds to $H=(1+\alpha)/2=1.52$. It means that the return time series has $H=0.52$. \cite{Serletis-Andreadis-2004-EE} also calculated the structure function of the price time series and found that $H=0.69$.

\cite{Serletis-Rosenberg-2007-PA} performed detrending moving average (DMA) analysis on the (logarithmic) WTI crude oil futures prices over the period from July 2, 1990 to November 1, 2006 and reported that $H\approx0.50$. This result indicates nicely that, on average, the WTI market was efficient in the weak form over the sample period. We also notice that the DMA fluctuation function has a better power-law scaling than most other Hurst index estimators reviewed in this work.

\cite{Elder-Serletis-2008-RFE} adopted a semi-parametric wavelet-based estimator to estimate the fractional integration parameter $d$ of the WTI oil prices sampled over the period from January 3, 1994 to June 30, 2005. They found that the fractional integration parameter was $d=-0.162$ when the Daubechies-4 wavelet was adopted and $d=-0.243$ when the Daubechies-6 wavelet was used. Since $H=0.5+d$, these results indicate that the WTI oil prices were anti-persistent over the sample period.

\cite{AlvarezRamirez-Alvarez-Rodriguez-2008-EE} applied the detrended fluctuation analysis (DFA) to WTI crude oil prices for the sample period 1987-2007. The DFA fluctuation function exhibited a crossover at $s_{\times}=25$ business days separating two power laws with very nice scaling behaviors. They found that the DFA scaling $H = 0.617\pm0.005$ for the short time horizons and $H = 0.507\pm0.004$ for the long time horizons. Because the DFA fluctuation functions of fractional Brownian motions always exhibit downward curvatures at small scales (the smaller the Hurst index, the larger the curvature) \citep{Bryce-Sprague-2012-SR,Shao-Gu-Jiang-Zhou-Sornette-2012-SR}, we argue that the scaling relation with $H = 0.507\pm0.004$ at large scales reflects the intrinsic behavior of the oil prices. It is intriguing to point out that this result is almost identical to that of \cite{Serletis-Rosenberg-2007-PA}. In contrast, with a longer sample from 1986 to 2009, \cite{AlvarezRamirez-Alvarez-Solis-2010-EE} reported a different DFA fluctuation function that could not be fitted by two power laws; rather, the function showed multiscaling behaviors.

\cite{Charles-Darne-2009-EP} conducted the variance ratio test on the WTI and Brent oil markets over the period 1982-2008. They divided the whole samples into two sub-periods at the end of 1993\footnote{The separation of the sample at the end of 1993 is rational: In 1993, in order to increase the efficiency of the North American energy industry, the United States,Canada and Mexico signed the North American Free Trade Agreement underpinning the process of deregulation \citep{Serletis-Andreadis-2004-EE,Serletis-RangelRuiz-2004-EE}}. They found that the Brent market were efficient on both sub-periods, while the WTI market was efficient on the 1982-1993 sub-period but inefficient on the 1994-2008 sub-period.

\cite{Gu-Chen-Wang-2010-PA} also adopted the detrended fluctuation analysis to the WTI and Brent oil prices sampled from 20 May 1987 to 30 September 2008. They also observed a crossover in each DFA fluctuation function at $s_{\times}=28$ for both markets. They found that $H=0.53$ when $s<s_{\times}$ and $H=0.45$ when $s>s_{\times}$ for WTI and $H=0.54$ when $s<s_{\times}$ and $H=0.47$ when $s>s_{\times}$. We argue that these results hardly support the inefficiency conjecture and roughly consistent with those of \cite{AlvarezRamirez-Alvarez-Rodriguez-2008-EE}. Further, \cite{Gu-Chen-Wang-2010-PA} divided each sample into three time periods separated by the middle of the Gulf War (February 24, 1991) and the Iraq War (March 30, 2003) and performed DFA on each time period. Crossover phenomena were also observed with $s_{\times}=21$ for the first period and $s_{\times}=20$ for the second and third periods. The Hurst indexes were found to vary in $[0.52,0.64]$ when $s<s_{\times}$ and in $[0.36,0.49]$ when $s>s_{\times}$.

\cite{Cunado-GilAlana-PerezDeGracia-2010-JFutM} utilized different methods to estimate the fractional integration parameter $d$ of WTI crude oil prices over the period from April 4, 1983 to September 9, 2008, including the modified R/S analysis of \cite{Lo-1991-Em}, the rescaled variance (V/S) analysis of \cite{Giraitis-Kokoszka-Leipus-Teyssiere-2003-JEm}, and the Whittle semiparametric method of \cite{Robinson-1995-AS}. Their results that the hypothesis $d=0$ could not be rejected at the 95\% confidence level indicated that there was little or no evidence of long memory in crude oil prices.

\cite{Wang-Wei-Wu-2011-PA} analyzed the WTI crude oil prices from January 2, 1990 to March 9, 2010 using the detrended fluctuation analysis and identified three scaling regimes. They found that $H=0.546$ when $s < 23$, $H=0.472$ when $23 < s < 250$ and $H=0.516$ when $s > 250$. We argue that their results are consistent with those of \cite{AlvarezRamirez-Alvarez-Rodriguez-2008-EE} and \cite{Gu-Chen-Wang-2010-PA}.

\cite{He-Qian-2012-PA} studied the spot prices of the WTI and Brent crude oil using the R/S analysis, the modified R/S analysis and the V/S analysis of \cite{Cajueiro-Tabak-2005-MCS} based on the work of \cite{Giraitis-Kokoszka-Leipus-Teyssiere-2003-JEm}. They estimated that $H=0.5406$ with the R/S analysis, $H=0.5166$ with the modified R/S analysis and $H=0.4731$ with the V/S analysis, which are respectively very close to the corresponding Hurst indexes 0.5308, 0.5122 and 0.4806 of the shuffled time series. It means that there is no long memory.

\subsection{Evolution of oil market efficiency in moving windows}

Following the idea of estimating the Hurst indexes of stock markets in moving windows of four years \citep{Cajueiro-Tabak-2004a-PA}, \cite{Tabak-Cajueiro-2007-EE} performed the classic R/S analysis on the daily Brent and WTI crude oil futures prices over the sample period 1983-2004 and found that both markets became more efficient and the WTI market was more efficient than the Brent market.

\cite{AlvarezRamirez-Alvarez-Rodriguez-2008-EE} adopted the detrended fluctuation analysis to study the WTI and Brent oil prices in moving windows. When the size of the moving window is chosen to be 175 business days, no  evident trend was found in the evolving Hurst index, which  oscillated around $H=0.617$. When longer moving windows with size of 300 business days were adopted, they found that the Hurst indexes decreased from 0.617 to 0.50 and the markets became efficient.

\cite{Charles-Darne-2009-EP} investigated the market efficiency of WTI and Brent oil (1982-2008) using the variance ratio tests. The time period was divided into two parts. Surprisingly, the WTI market turned from weak-form efficiency to inefficiency, indicating that the deregulation of NATFA did not improve the efficiency on the WTI crude oil market as desired.

\cite{Wang-Liu-2010-EE} tested the dynamic efficiency of the WTI crude oil market over sample period from July 13, 1990 to March 6, 2009. The local Hurst indexes were estimated in moving windows using multiscale detrended fluctuation analysis. They found that the short-term, medium-term and long-term behaviors were generally turning into more efficient over time.

Equipped with the R/S, V/S and modified R/S analyses, \cite{He-Qian-2012-PA} studied the dynamic efficiency of the WTI and Brent crude oil spot markets in moving windows. They found that the three Hurst index functions $H_{RS}$, $H_{MRS}$ and $H_{VS}$ have similar U-shaped evolution patters and $H_{RS}>H_{MRS}>H_{VS}$ at each time. Based on numerical experiments, \cite{He-Qian-2012-PA} argued that the R/S analysis overestimates the Hurst index, the V/S analysis underestimates the Hurst index, and the modified R/S analysis provides better estimates. In this sense, we argue that both the Brent and WTI spot markets seem to be efficient since the Hurst index evolved within the interval $[0.45,0.55]$, which is expected to be not significantly different from 0.5 \citep{Weron-2002-PA}.

\subsection{Discussions}

Our brief review has unfolded the evolution of the topic as well as the conclusions. Many different methods have been adopted, partially owning to the fact that our understanding of the performances of different estimators also evolved. It is also not surprising that a same method may result in different estimates of the Hurst index, because different sample periods and different scaling ranges have been chosen.

There are more than ten techniques that have been invented to detect long-range correlations in time series \citep{Taqqu-Teverovsky-Willinger-1995-Fractals,Delignieres-Ramdani-Lemoine-Torre-Fortes-Ninot-2006-JMPsy,Kantelhardt-2009-ECSS}, such as the rescaled range (R/S) analysis \citep{Hurst-1951-TASCE}, the wavelet transform module maxima approach \citep{Holschneider-1988-JSP,Muzy-Bacry-Arneodo-1991-PRL,Bacry-Muzy-Arneodo-1993-JSP}, the fluctuation analysis (FA) \citep{Peng-Buldyrev-Goldberger-Havlin-Sciortino-Simons-Stanley-1992-Nature}, the detrended fluctuation analysis (DFA) \citep{Peng-Buldyrev-Havlin-Simons-Stanley-Goldberger-1994-PRE}, the detrending moving average analysis (DMA) \citep{Alessio-Carbone-Castelli-Frappietro-2002-EPJB}, and their variants. There is no consensus which estimators perform best. However, DFA and DMA are argued to be ``The Methods of Choice'' in determining the Hurst index of time series \citep{Shao-Gu-Jiang-Zhou-Sornette-2012-SR}.

One possible way to solve the problem is to perform statistical tests. For a given estimator, the Hurst index $H$ of the original time series is determined. Then, one can shuffle the time series many times and estimate the Hurst index for each shuffled time series. The Hurst index $H$ is compared with the average Hurst index $\langle{H}\rangle$ of the shuffled time series. If the null hypothesis $H=\langle{H}\rangle$ cannot be rejected, the original time series can be argued to possess no long-term correlations. Although there are several well conducted studies \citep{Weron-2002-PA,Couillard-Davison-2005-PA}, this approach has not been adopted in the majority of previous works.

\section{Data description}

\begin{figure}[htb]
  \centering
  \includegraphics[width=4.3cm]{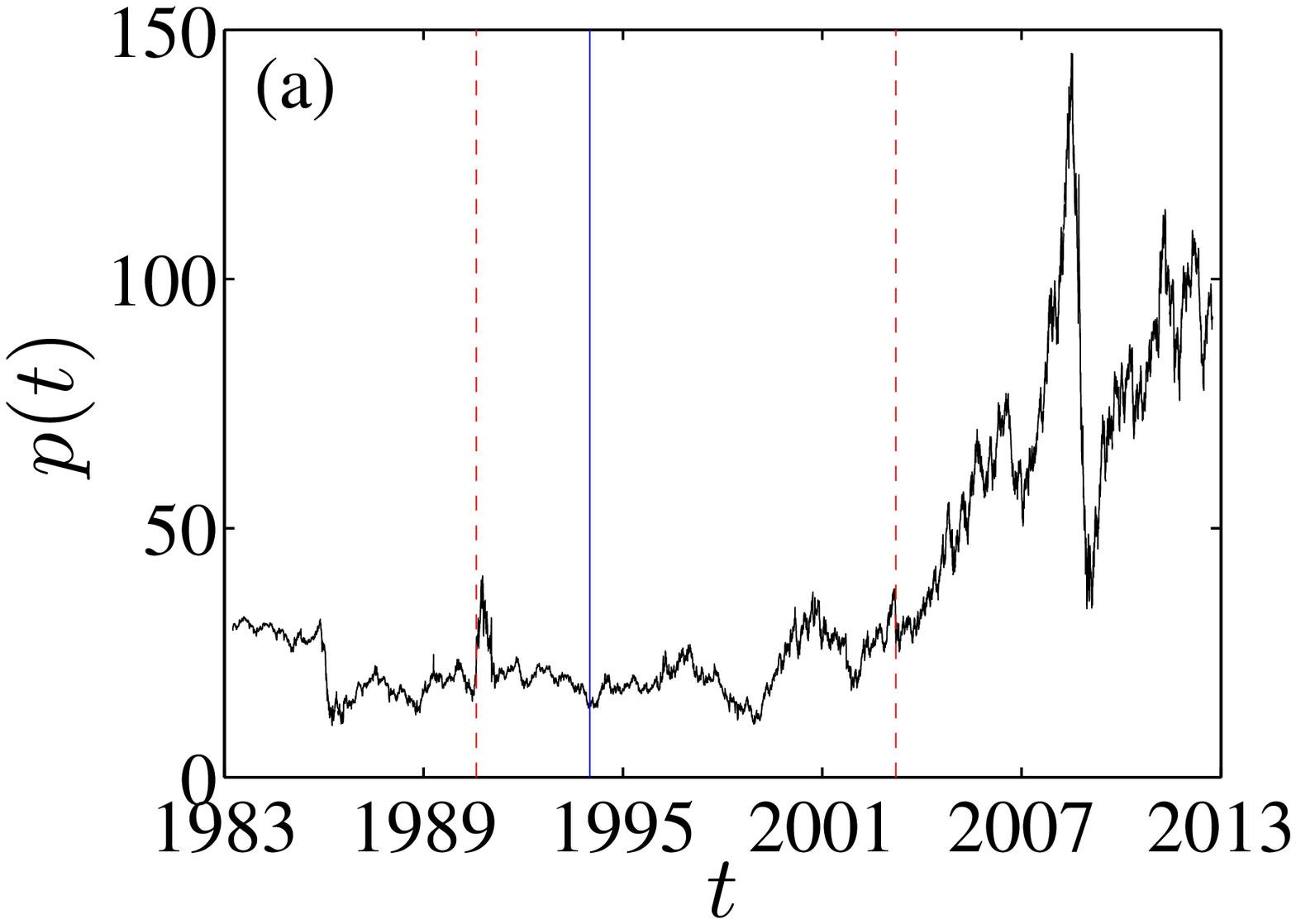}
  \includegraphics[width=4.3cm]{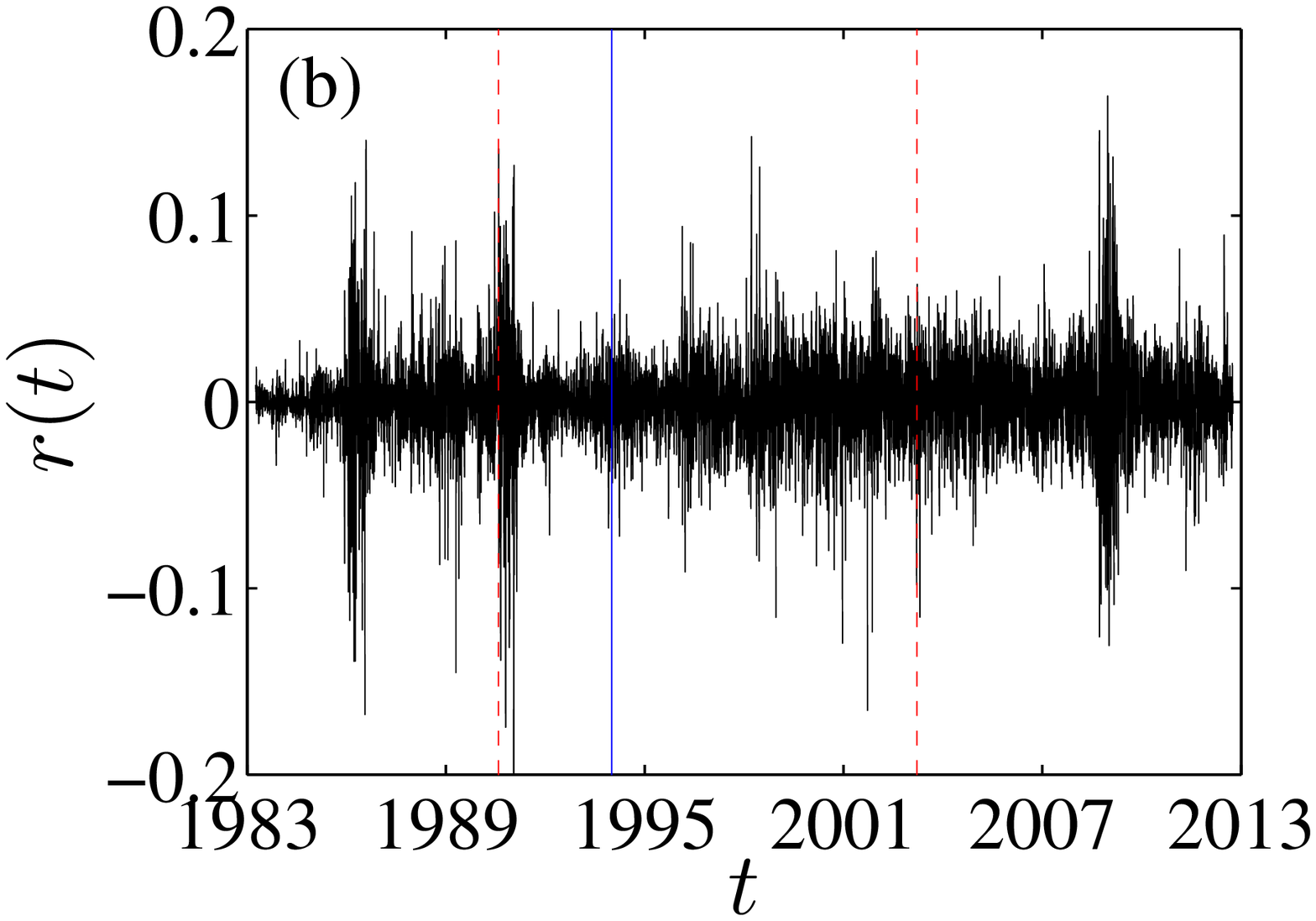}
  \caption{(Color online.) (a) Evolution of the daily prices of WTI crude oil futures. (b) Evolution of the returns of WTI crude oil futures. The vertical lines mark the outbreak of the Gulf War on August 2, 1990, the signing date of the NAFTA on January 1, 1994, and the outbreak of the Iraq War on March 20, 2003.}
  \label{Fig:Date:WTI:Price:Return}
\end{figure}

Our analysis is based on the daily prices of the WTI crude oil futures, which are freely available at the web site of the U.S. Energy Information Administration ({\url{http://www.eia.gov/}}). Our data cover the period from April 4, 1983 to October 2, 2012. There are totally 7401 data points. Figure \ref{Fig:Date:WTI:Price:Return}(a) plots the evolution of the daily prices of WTI crude oil futures. The outbreaks of two wars, the Gulf War broken out on August 2, 1990 and the Iraq War broken out on March 20, 2003, are shown as two dashed vertical lines. The solid vertical line corresponds to the date January 1, 1994, on which the North American Free Trade Agreement are signed by three countries, the United States, Canada, and Mexico.

The daily returns are defined as
\begin{equation}
  r(t) = \ln{P(t)}-\ln{P(t-1)}
\end{equation}
The evolution of the returns of WTI crude oil futures is illustrated in Fig.~\ref{Fig:Date:WTI:Price:Return}(b). One can observe that there are many large returns after the outbreak of the Gulf war. In sharp contrast to this, the fluctuating behavior of the returns almost keeps the same as that of the returns before the outbreak of the Iraq war. After signing the North American Free Trade Agreement, one can observe that the fluctuations of the returns became gradually smaller in the following two years.

\section{Methods}

The detrended fluctuation analysis (DFA) was invented by \cite{Peng-Buldyrev-Havlin-Simons-Stanley-Goldberger-1994-PRE} originally to investigate the long-range dependence in coding and noncoding DNA nucleotide sequences. The properties of DFA have been extensively studied \citep{Hu-Ivanov-Chen-Carpena-Stanley-2001-PRE,Chen-Ivanov-Hu-Stanley-2002-PRE,Chen-Hu-Carpena-BernaolaGalvan-Stanley-Ivanov-2005-PRE,Shang-Lin-Liu-2009-PA,Xu-Shang-Kamae-2009-CSF,Ma-Bartsch-BernaolaGalvan-Yoneyama-Ivanov-2010-PRE}. The detrending moving average (DMA) analysis is based on the moving average or mobile average technique \citep{Carbone-2009-IEEE}, which was first proposed by \cite{Vandewalle-Ausloos-1998-PRE} to estimate the Hurst exponent of self-affinity signals and further developed by \cite{Alessio-Carbone-Castelli-Frappietro-2002-EPJB} by considering the second-order difference between the original signal and its moving average function. The DMA method has been widely applied to the analysis of real-world time series \citep{Carbone-Castelli-2003-SPIE,Carbone-Castelli-Stanley-2004-PA,Carbone-Castelli-Stanley-2004-PRE,Varotsos-Sarlis-Tanaka-Skordas-2005-PRE,Serletis-Rosenberg-2007-PA,Arianos-Carbone-2007-PA,Matsushita-Gleria-Figueiredo-Silva-2007-PLA,Serletis-Rosenberg-2009-CSF} and synthetic signals \citep{Carbone-Stanley-2004-PA,Xu-Ivanov-Hu-Chen-Carbone-Stanley-2005-PRE,Serletis-2008-CSF,Gu-Zhou-2010-PRE}. Extensive numerical experiments unveil that the performance of the DMA method are comparable to the DFA method with slightly different priorities under different situations \citep{Xu-Ivanov-Hu-Chen-Carbone-Stanley-2005-PRE,Bashan-Bartsch-Kantelhardt-Havlin-2008-PA,Shao-Gu-Jiang-Zhou-Sornette-2012-SR}.

Both methods can be extended, for instance, to investigate multifractal time series \citep{CastroESilva-Moreira-1997-PA,Weber-Talkner-2001-JGR,Kantelhardt-Zschiegner-KoscielnyBunde-Havlin-Bunde-Stanley-2002-PA,Gu-Zhou-2010-PRE}, high-dimensional fractal and multifractal objects \citep{Gu-Zhou-2006-PRE,Carbone-2007-PRE,Turk-Carbone-Chiaia-2010-PRE}, and long-term power-law cross-correlations \citep{Podobnik-Stanley-2008-PRL,Zhou-2008-PRE,Podobnik-Horvatic-Petersen-Stanley-2009-PNAS,Jiang-Zhou-2011-PRE}.

The DFA and DMA share the same framework as described below. In the first step, the cumulative summation series $y_i$ is determined as follows
\begin{equation}
  y_i = \sum_{j=1}^{i} \left(r_i-\langle{r}\rangle\right),~~i = 1, 2, \cdots, N,
  \label{Eq:cumsum}
\end{equation}
where $\langle{r}\rangle$ is the sample mean of the return series.

The main difference between the DFA and DMA algorithms is the determination of the ``local trend function'' $\widetilde{y}_i$, which is dependent of the box size $s$. In the DFA algorithm, the local trend $\widetilde{y}_i$ is determined by polynomial fits in boxes \citep{Peng-Buldyrev-Havlin-Simons-Stanley-Goldberger-1994-PRE}. In the DMA approach, one calculates the moving average function $\widetilde{y}_i$ in a moving window \citep{Arianos-Carbone-2007-PA},
\begin{equation}
  \widetilde{y}_i(s)=\frac{1}{s}\sum_{k=-\lfloor(s-1)\theta\rfloor}^{\lceil(s-1)(1-\theta)\rceil}y_{i-k},
  \label{Eq:1ddma:y1}
\end{equation}
where $s$ is the window size, $\lfloor{x}\rfloor$ is the largest integer not greater than $x$, $\lceil{x}\rceil$ is the smallest integer not smaller than $x$, and $\theta$ is the position parameter with the value varying in the range $[0,1]$. Hence, the moving average function considers $\lceil(s-1)(1-\theta)\rceil$ data points in the past and $\lfloor(s-1)\theta\rfloor$ points in the future. We consider three special cases in this paper. The first case $\theta=0$ refers to the backward moving average \citep{Xu-Ivanov-Hu-Chen-Carbone-Stanley-2005-PRE}, in which the moving average function $\widetilde{y}_i$ is calculated over all the past $s-1$ data points of the signal. The second case $\theta=0.5$ corresponds to the centered moving average \citep{Xu-Ivanov-Hu-Chen-Carbone-Stanley-2005-PRE}, where $\widetilde{y}_i$ contains half past and half future information in each window. The third case $\theta=1$ is called the forward moving average, where $\widetilde{y}_i$ considers the trend of $s-1$ data points in the future.

The residual sequence $\epsilon_i$ is obtained by removing the local trend function $\widetilde{y}_i$ from $y_i$:
\begin{equation}
  \epsilon_i(s)=y_i-\widetilde{y}_i(s).
  \label{Eq:1ddma:epsilon}
\end{equation}
We can then calculate the overall fluctuation function $F(s)$ as follows
\begin{equation}
  \left[F(s)\right]^2 = \frac{1}{N}\sum_{i=1}^{N} \left[\epsilon_i(s)\right]^2.
  \label{Eq:F2:s}
\end{equation}
As the box size $s$ varies in proper scaling ranges, one can determine the power law relationship between the overall fluctuation function $F(s)$ and the box size $s$,
\begin{equation}
  F(s) \sim s^H,
  \label{Eq:Hurst}
\end{equation}
where $H$ signifies the DFA or DMA scaling exponent.

It is necessary to emphasize that the scaling exponent $H$ obtained from Eq.~(\ref{Eq:Hurst}) should be called ``DFA exponent'' or ``DMA exponent'' rather than ``Hurst index'' in the rigorous sense. For fractional Brownian motions (fBms), the DFA/DMA exponent is identical to the Hurst index, which is related to the power spectrum exponent $\eta$ by $\eta = 2H-1$ and to the autocorrelation exponent $\gamma$ by $\gamma = 2-2H$ \citep{Talkner-Weber-2000-PRE,Heneghan-McDarby-2000-PRE,Arianos-Carbone-2007-PA}. In addition, \cite{Serinaldi-2010-PA} provided a nice clarification on this issue. If the Hurst index of a fractional Gaussian noise is $H$, then the Hurst index of its associated fractional Brownian motion is also $H$. If one replaces $P_i$ for $r_i$ in the cumulative summation procedure (\ref{Eq:cumsum}), the resultant scaling exponent cannot be termed Hurst index. In this case, many estimators give a scaling exponent larger than 1.

\section{Results}
\label{S1:Results}

In this section, we provide the analysis results of the memory behaviors in the WTI crude oil futures prices by means of DMA and DFA approaches. The investigated sample includes the whole series, five sub-series, and local samples in moving windows. We perform bootstrapping-based statistical tests to access if a time series under investigation possesses long-term correlations.

\subsection{The whole sample}

\begin{figure}[htb]
  \centering
  \includegraphics[width=4.3cm]{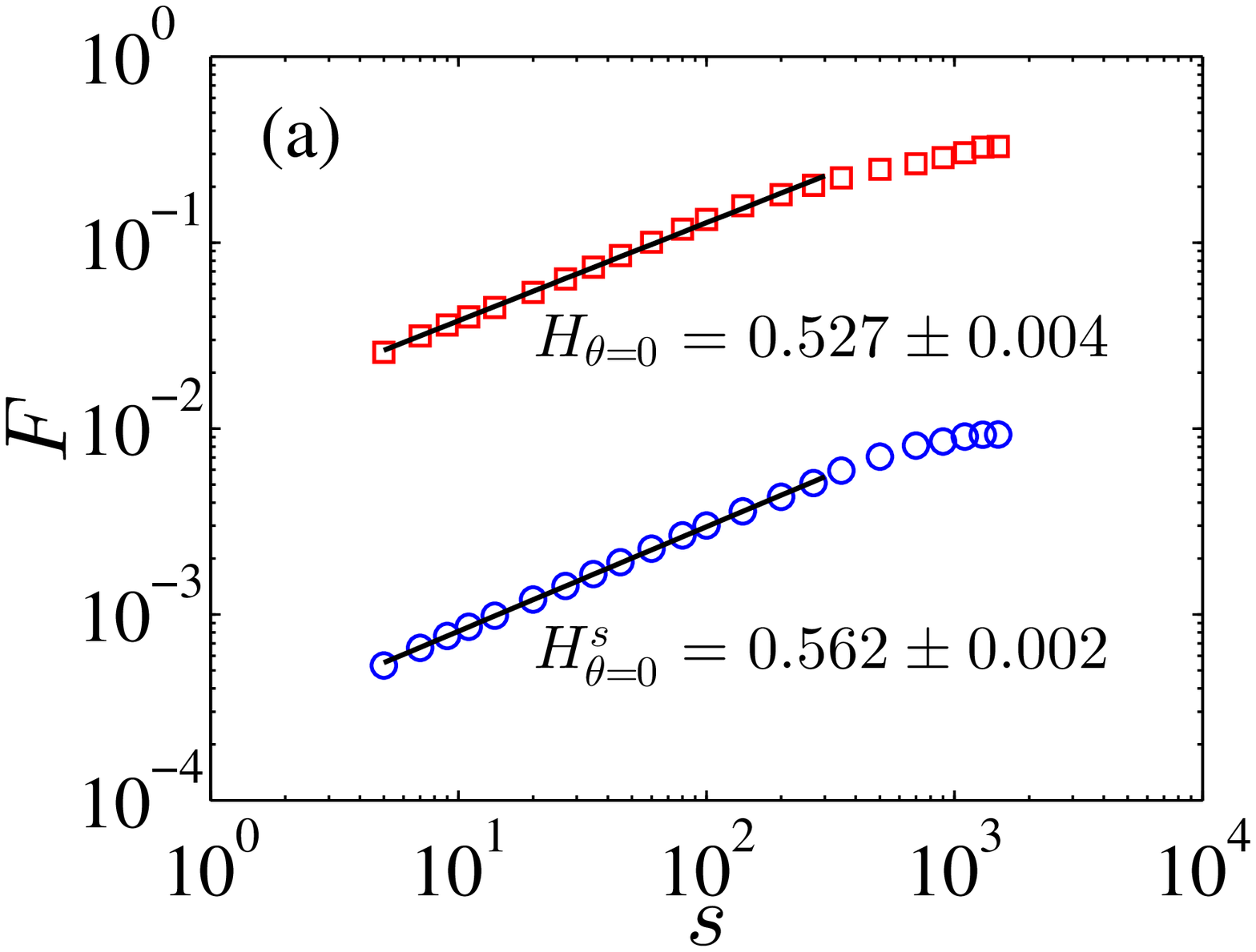}
  \includegraphics[width=4.3cm]{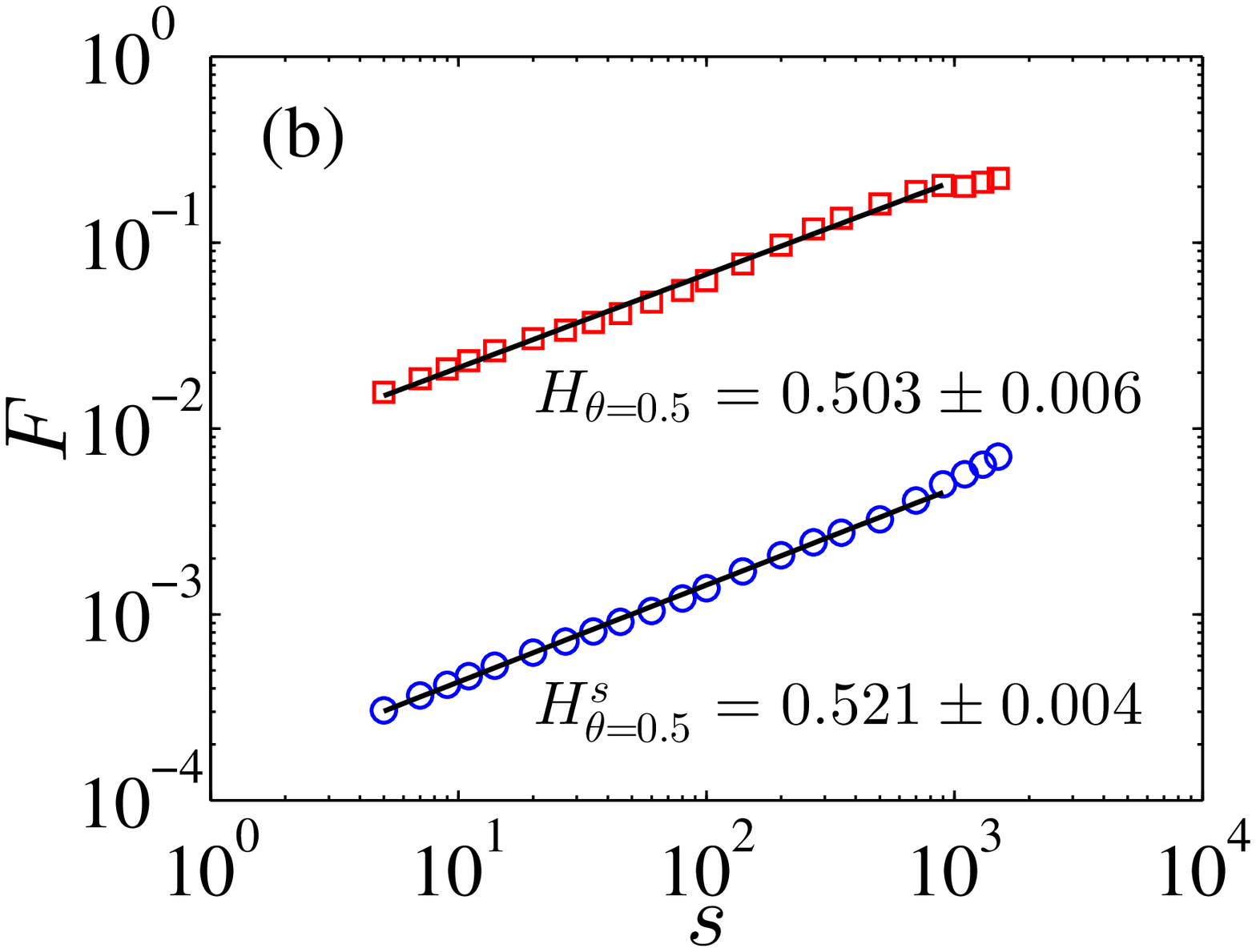}
  \includegraphics[width=4.3cm]{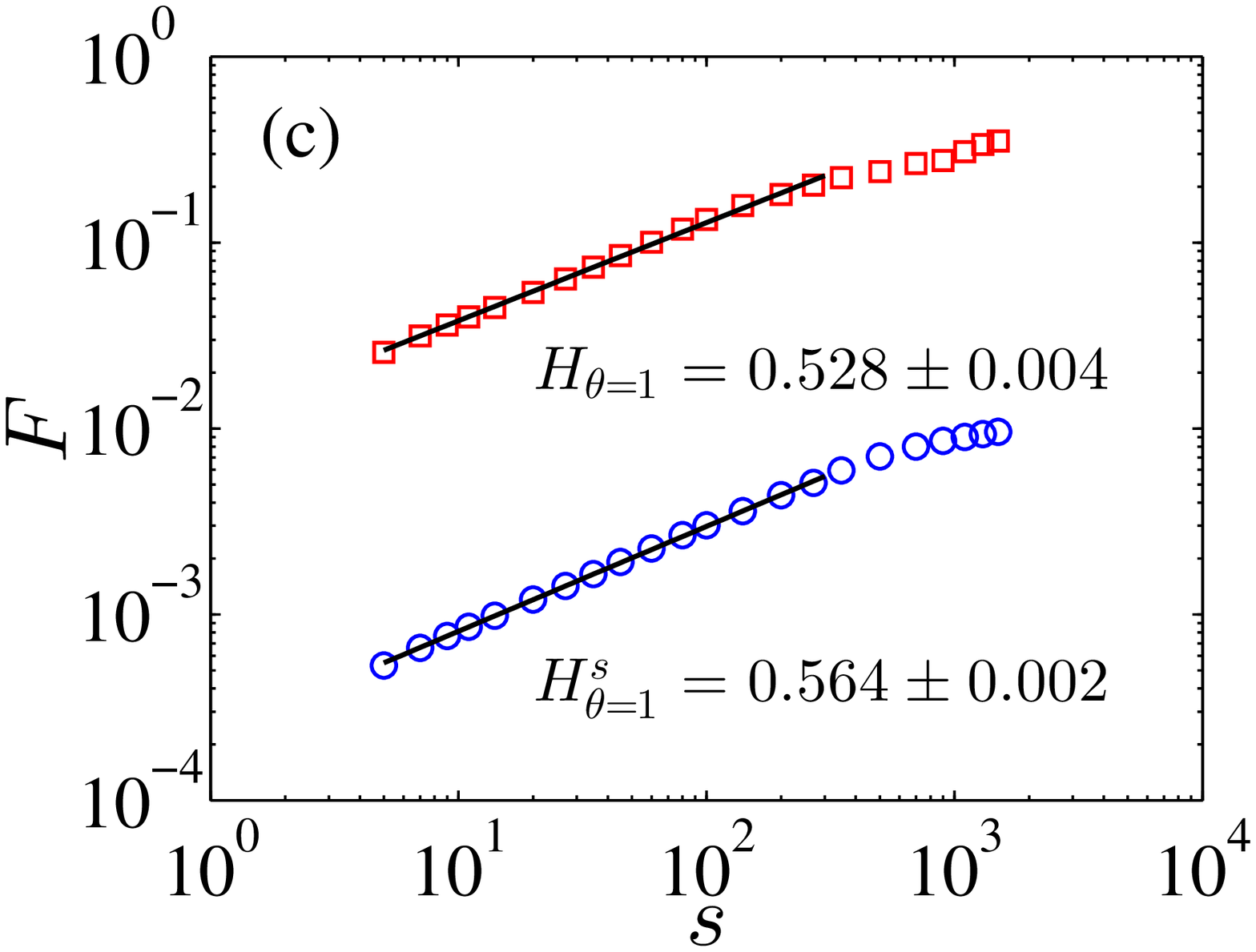}
  \includegraphics[width=4.3cm]{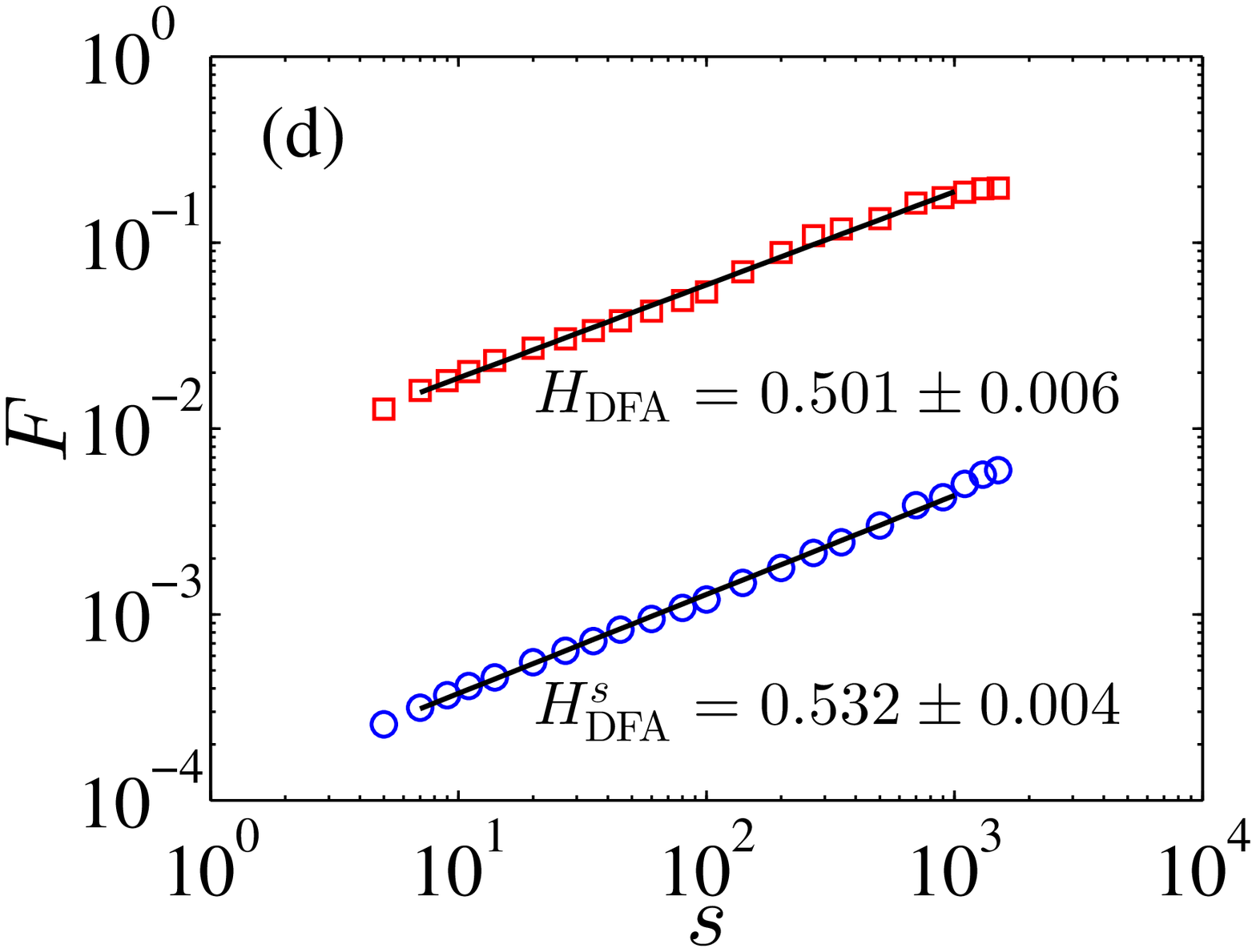}
  \caption{(Color online.) Plots of the fluctuation function $F$ with respect to the box size $s$ for the original series and shuffled series. (a) DMA with $\theta = 0$. (b) DMA with $\theta = 0.5$. (c) DMA with $\theta = 1$ (d) DFA.} \label{Fig:Fut:Fs}
\end{figure}

To have an overview of the memory behaviors in the WTI crude oil futures, we analyze the whole series by means of DMA and DFA approaches. Figure \ref{Fig:Fut:Fs} illustrates the fluctuation function $F$ with respect to the box size $s$ as open squares for DMA ($\theta = 0$ in panel (a), $\theta = 0.5$ in panel (b), and $\theta = 1$ in panel (c)) and DFA (panel (d)). One can observe very nice power-law scaling behaviors between the fluctuation function $F$ and the box size $s$ in each panel and the solid lines are the best power-law fits to the data points in corresponding scaling ranges, which gives the exponents $H = 0.527 \pm 0.004$ for DMA with $\theta = 0$, $H = 0.503 \pm 0.006$ for DMA with $\theta = 0.5$, $H = 0.528 \pm 0.004$ for DMA with $\theta = 1$, and $H = 0.501 \pm 0.006$ for DFA, respectively. 

For comparison, we also perform the same analysis on the shuffled series, which is obtained by reshuffling the original series to remove possible memory behaviors. The purpose of the comparison is to check whether the original series exhibit the same memory behaviors as the shuffled series. The consequence for a randomly chosen realization of the shuffled time series is plotted as open circles in each panel as well. One can find that the estimated exponents of the shuffled series are comparable to those of the original series, providing an impression that the return series of the WTI crude oil futures may exhibit same memory features as its shuffled series.

To distinguish the memory behaviors in original series and shuffled series through a strict statistical way, we propose a statistical test in the spirit of bootstrapping. The statistic is defines as the DMA or DFA exponent $H$. In order to obtain an ensemble of the statistic $H$, which allows us to generate a distribution of the statistic, we reshuffle the original series for 10,000 times and estimate the DMA or DFA exponent $H^s$ for each shuffled series. The null hypothesis is that the original series possesses the same memory traits as the shuffled series, {\it{i.e.}}, $H=\langle{H^s}\rangle$, where $\langle{H^s}\rangle$ is the mean of all $H^s$ values. 

\begin{figure}[htb]
  \centering
  \includegraphics[width=4.3cm]{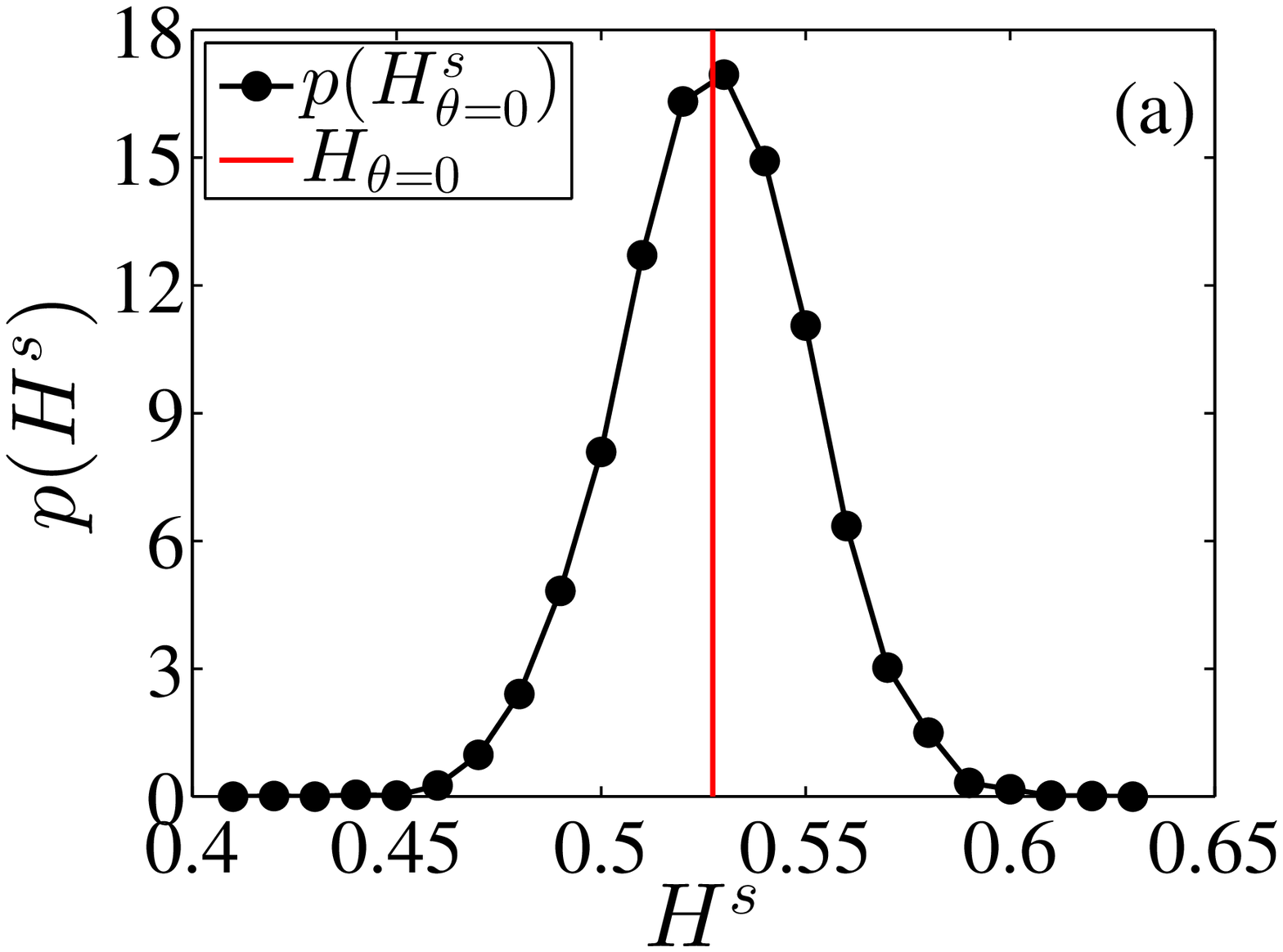}
  \includegraphics[width=4.3cm]{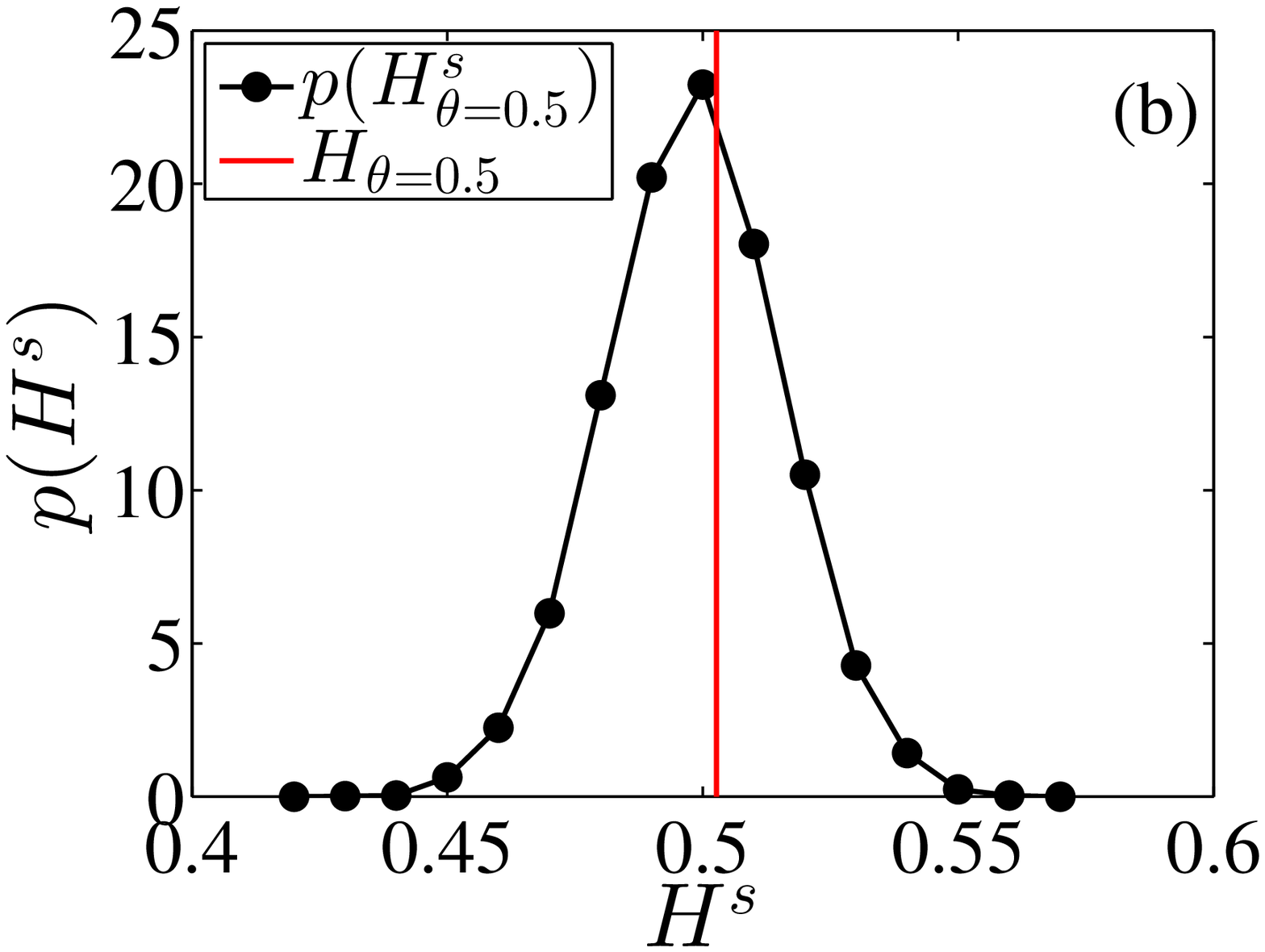}
  \includegraphics[width=4.3cm]{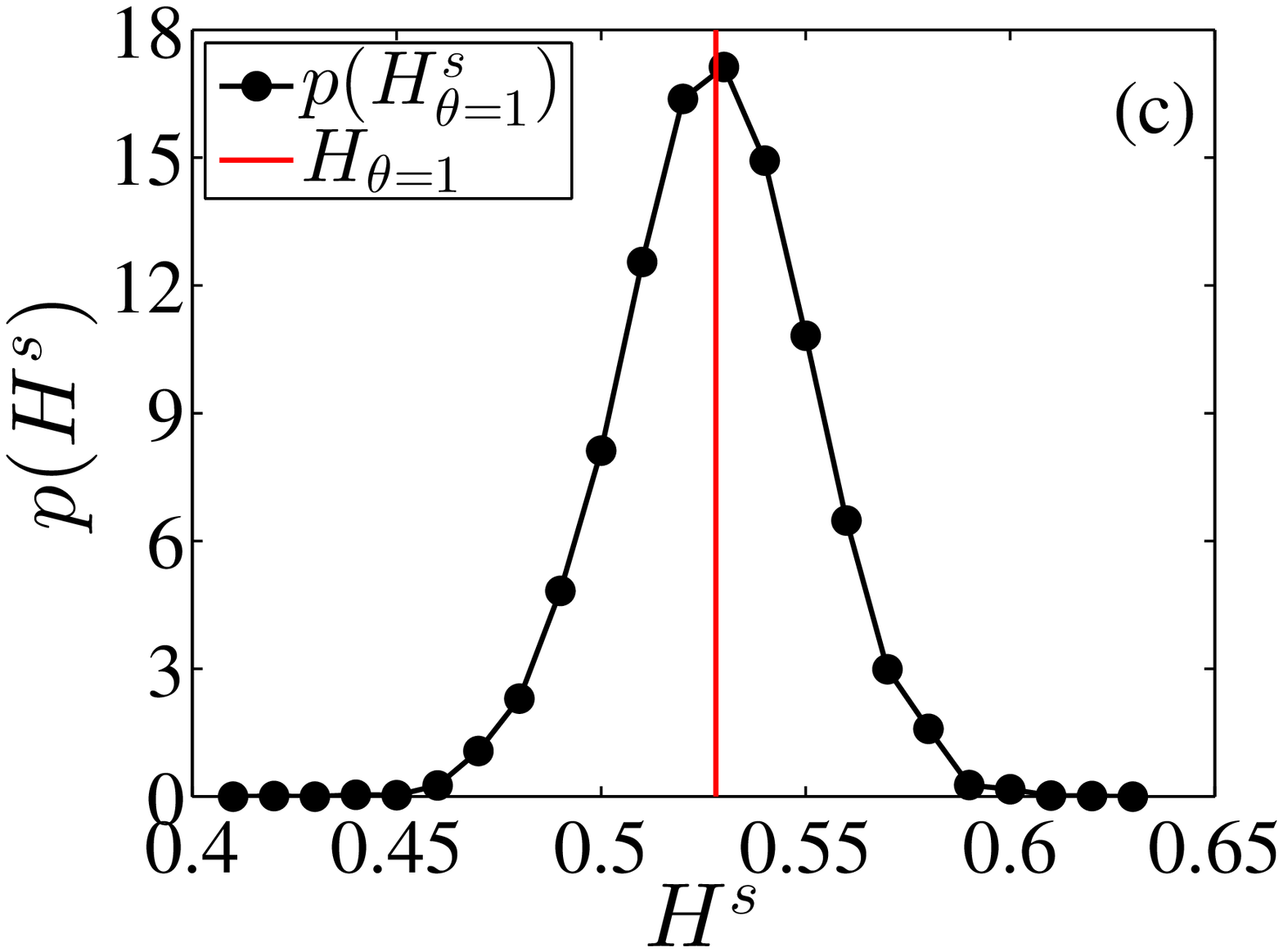}
  \includegraphics[width=4.3cm]{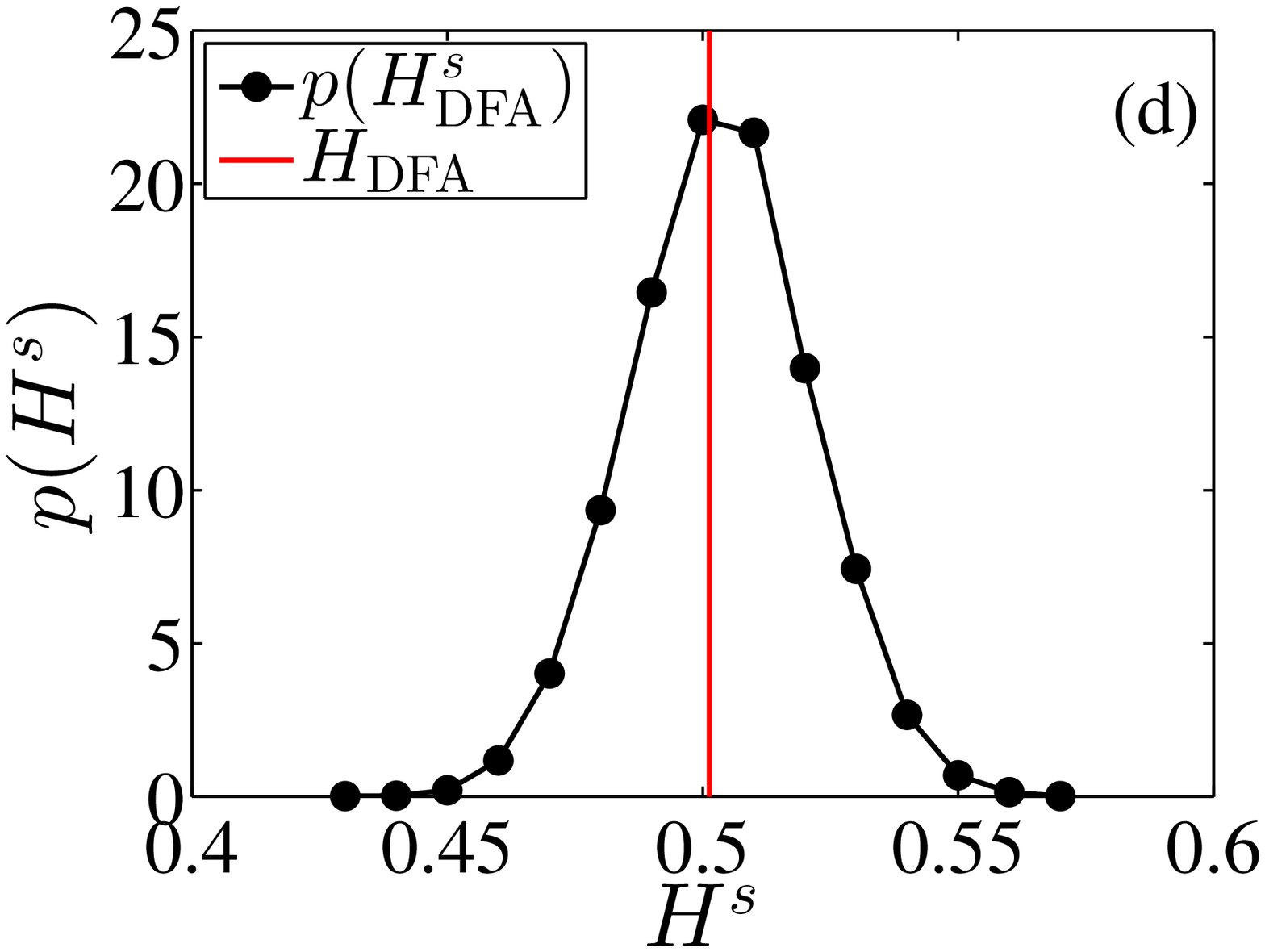}
  \caption{(Color online.) Probability distributions of the DMA and DFA exponent $H^s$ obtained from the shuffled series. The vertical lines correspond to the DMA or DFA exponent $H$ estimated from the original series. (a) DMA with $\theta = 0$. (b) DMA with $\theta = 0.5$. (c) DMA with $\theta = 1$ (d) DFA.}
  \label{Fig:Fut:PDF:H}
\end{figure}

Figure \ref{Fig:Fut:PDF:H} illustrates the probability distribution of the estimated exponent $H^s$ of the shuffled series for the four approaches. The vertical line in each panel corresponds to the DMA or DFA exponent of the original series, which locates at the central part of the distribution.

The difference in memory behaviors between the original series and the shuffled series is quantitatively described by a two-tailed $p$-value, which is defined as
\begin{equation}
  p = {\mathrm{Prob}}\left(|H^s - \langle H^s \rangle| > |H - \langle H^s \rangle|\right).
  \label{Eq:pvalue}
\end{equation}
We find that the $p$-value is 0.998 for DMA with $\theta = 0$, 0.793 for DMA with $\theta = 0.5$, 0.9757 for DMA with $\theta = 1$, and 0.902 for DFA, respectively. The results are presented in Table~\ref{Tb:Results:SubSamples}. At the significance level of 0.01, we cannot reject the null hypothesis, which offers the strong evidence supporting no memory behaviors in return series of the WTI crude oil futures.

\setlength\tabcolsep{3.5pt}
\begin{table}[htp]
\begin{center}
  \caption{\label{Tb:Results:SubSamples} Estimated DMA and DFA exponents of the whole series and the five sub-series and the results of bootstrapping tests. BDMA, CDMA and FDMA are respectively backward, centred and forward DMA.}
  \medskip
\begin{tabular}{l|cccccccc}
  \hline\hline
     && Whole & Sub 1 & Sub 2 & Sub 3 & Sub 4 & Sub 5 \\
  \hline
  BDMA &$H$                   & 0.527 & 0.554 & 0.463 & 0.590 & 0.571 & 0.530 \\
       &$\langle H^s \rangle$ & 0.527 & 0.502 & 0.502 & 0.516 & 0.525 & 0.522 \\
       &$p$                   & 0.998 & 0.341 & 0.394 & 0.102 & 0.204 & 0.778 \\
  \hline
  CDMA &$H$                   & 0.503 & 0.502 & 0.401 & 0.440 & 0.433 & 0.482 \\
       &$\langle H^s \rangle$ & 0.498 & 0.496 & 0.497 & 0.497 & 0.497 & 0.498 \\
       &$p$                   & 0.793 & 0.860 & {\bf 0.001} & 0.029 & {\bf 0.010} & 0.467 \\
  \hline
  FDMA &$H$                   & 0.528 & 0.557 & 0.452 & 0.591 & 0.564 & 0.528 \\
       &$\langle H^s \rangle$ & 0.527 & 0.502 & 0.510 & 0.516 & 0.520 & 0.520 \\
       &$p$                   & 0.976 & 0.318 & 0.158 & 0.098 & 0.257 & 0.823 \\
  \hline
  DFA  &$H$                   & 0.501 & 0.481 & 0.419 & 0.496 & 0.439 & 0.501 \\
       &$\langle H^s \rangle$ & 0.503 & 0.502 & 0.503 & 0.503 & 0.508 & 0.503 \\
       &$p$                   & 0.902 & 0.494 & {\bf 0.001} & 0.817 & {\bf 0.002} & 0.931 \\
  \hline\hline
\end{tabular}
\end{center}
\end{table}

From our analysis we can conclude that the crude oil market is efficient when the whole period (1983-2012) is taken into consideration.

\subsection{Several sub-samples}

To understand the influences of the important history events, which have great shocks on the crude oil market, on the memory behaviors in the WIT crude oil futures, we further apply the DMA and DFA approaches to five sub-series. Our purpose is to find out whether the exogenous events could change the efficiency of the oil market in the defined sub-periods. The five sub-series are determined as follows. We first divide the whole series into three sub-series with two separating dates. One is August 2, 1992, on which the Gulf War broke out, and the other is March 20, 2003, on which the Iraq War broke out. The three sub-series are denoted as sub-series 1, sub-series 2, and sub-series 3 based on their chronological orders. We also separate the whole series into two sub-series delimited at March 1, 1994, on which the North American Free Trade Agreement is signed by the United States, Canada and Mexico to increase the efficiency of the North American energy industry. The two sub-series are denoted as sub-series 4 and sub-series 5 according to their time sequence.

%

\begin{figure}[htb]
  \centering
  \includegraphics[width=4.3cm]{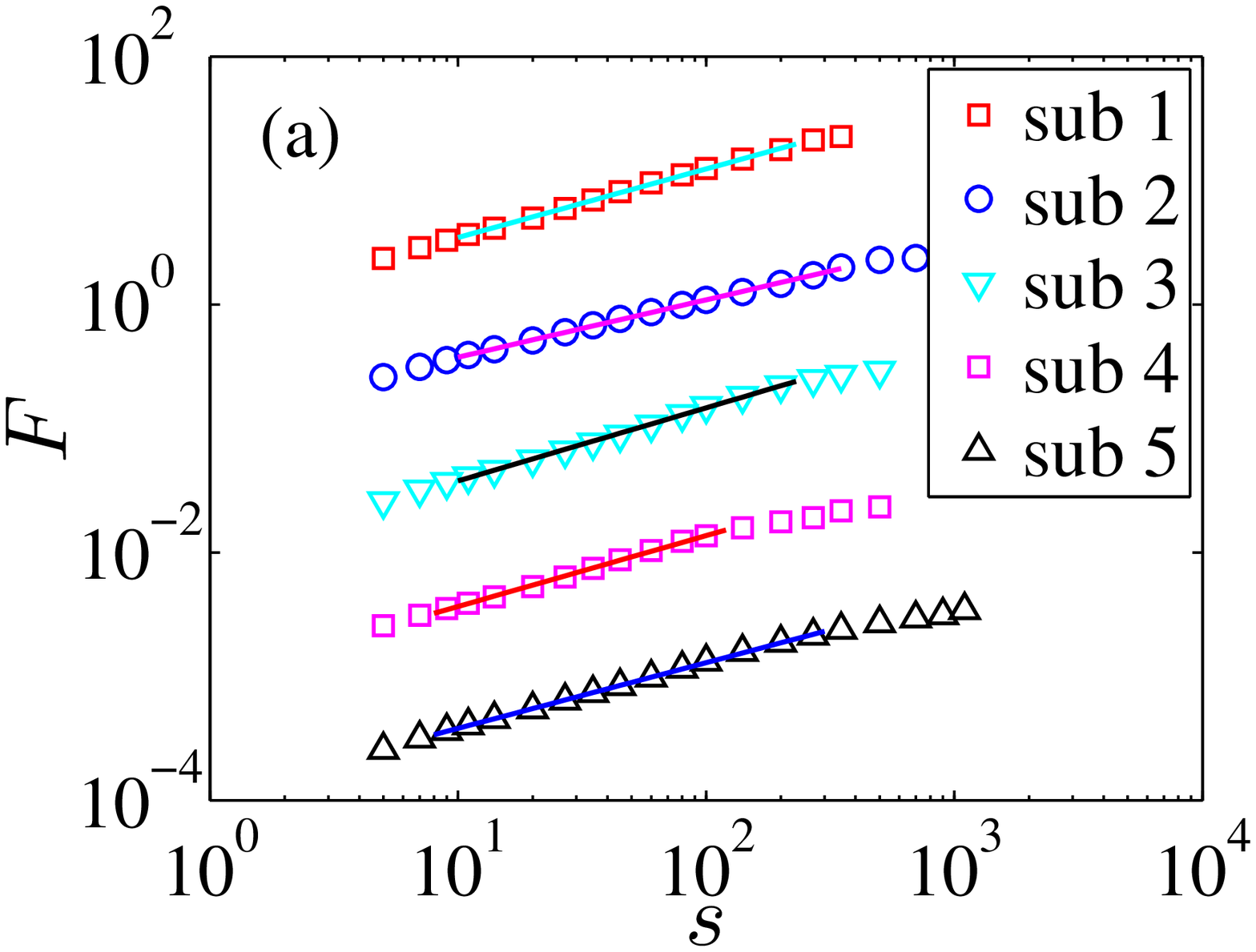}
  \includegraphics[width=4.3cm]{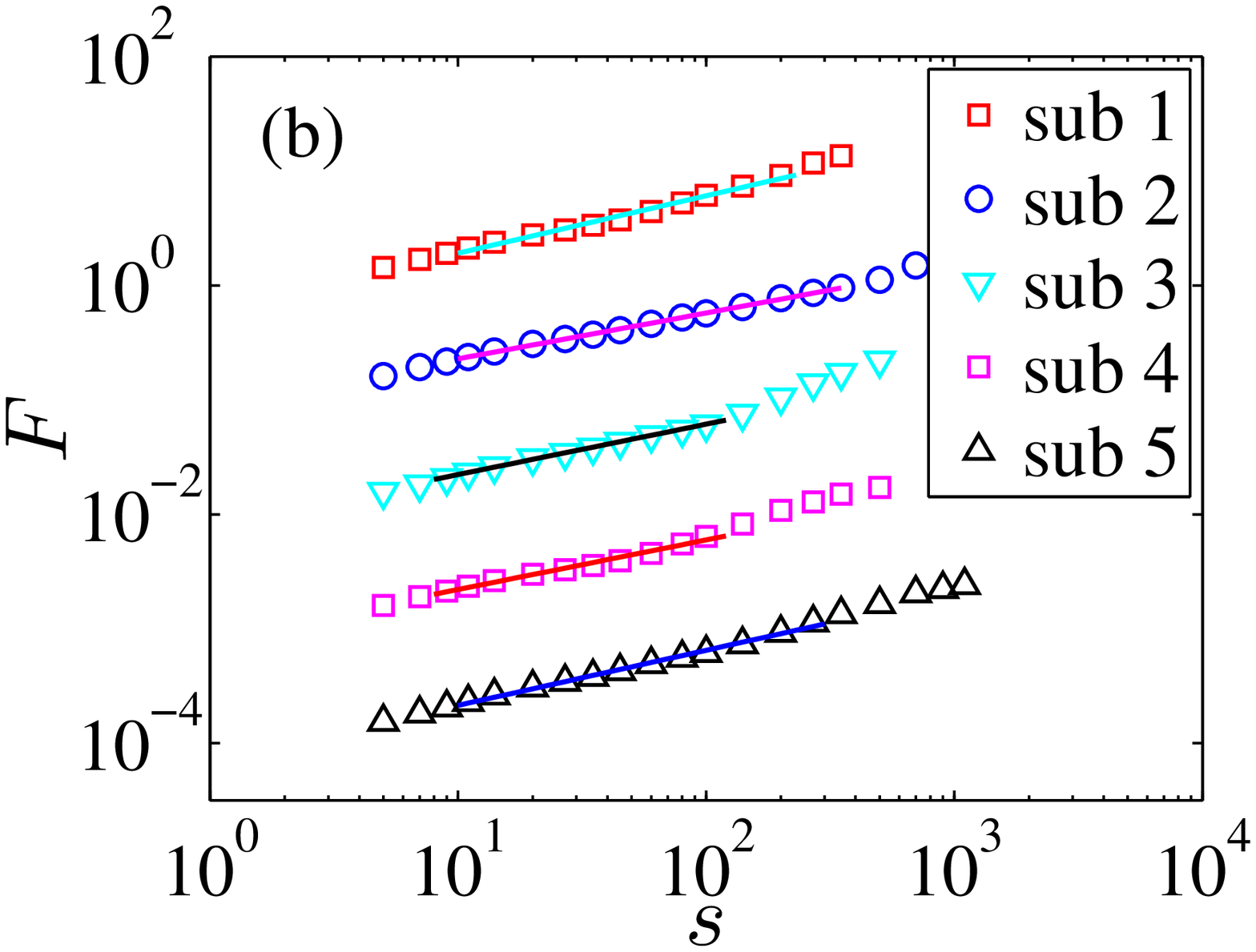}
  \includegraphics[width=4.3cm]{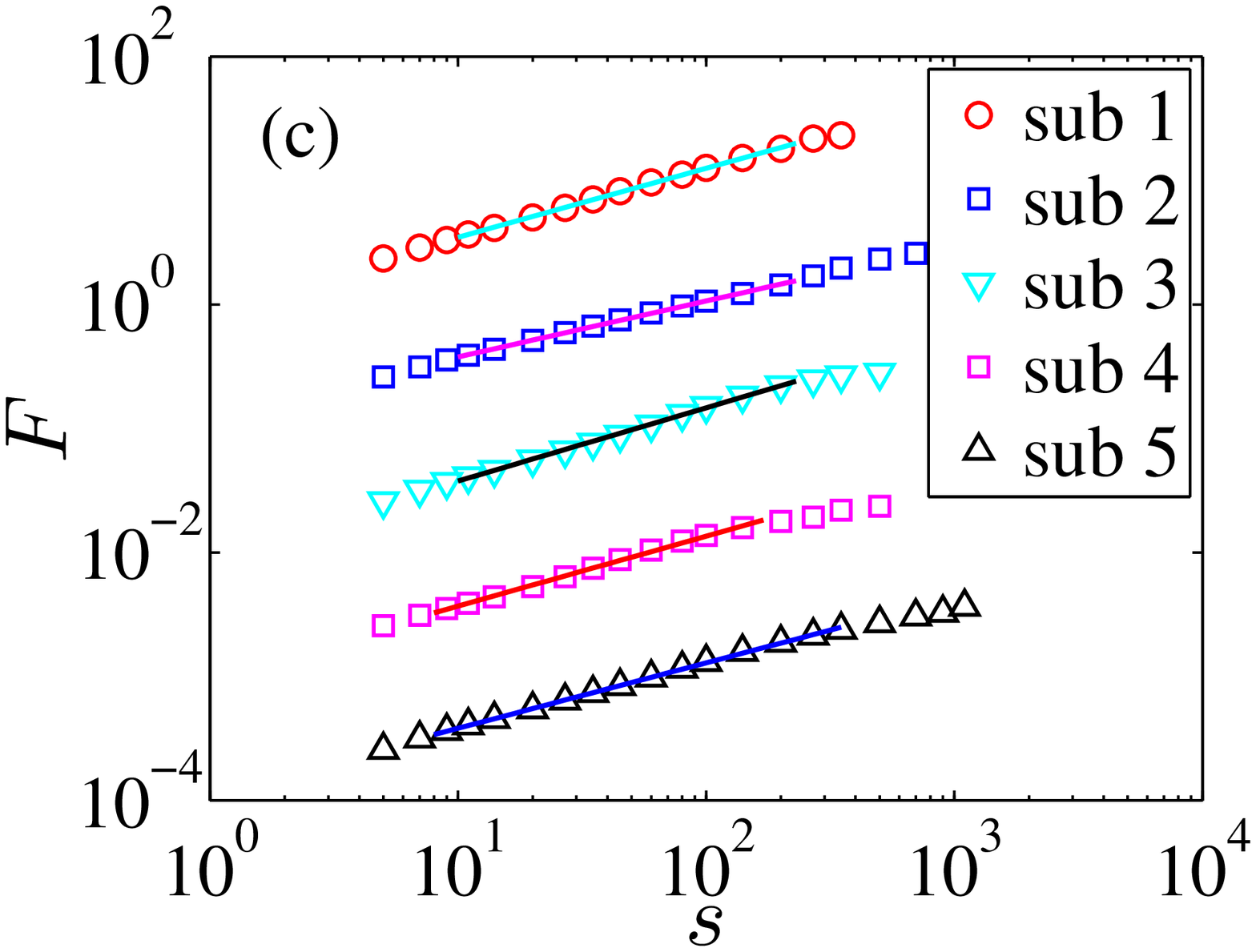}
  \includegraphics[width=4.3cm]{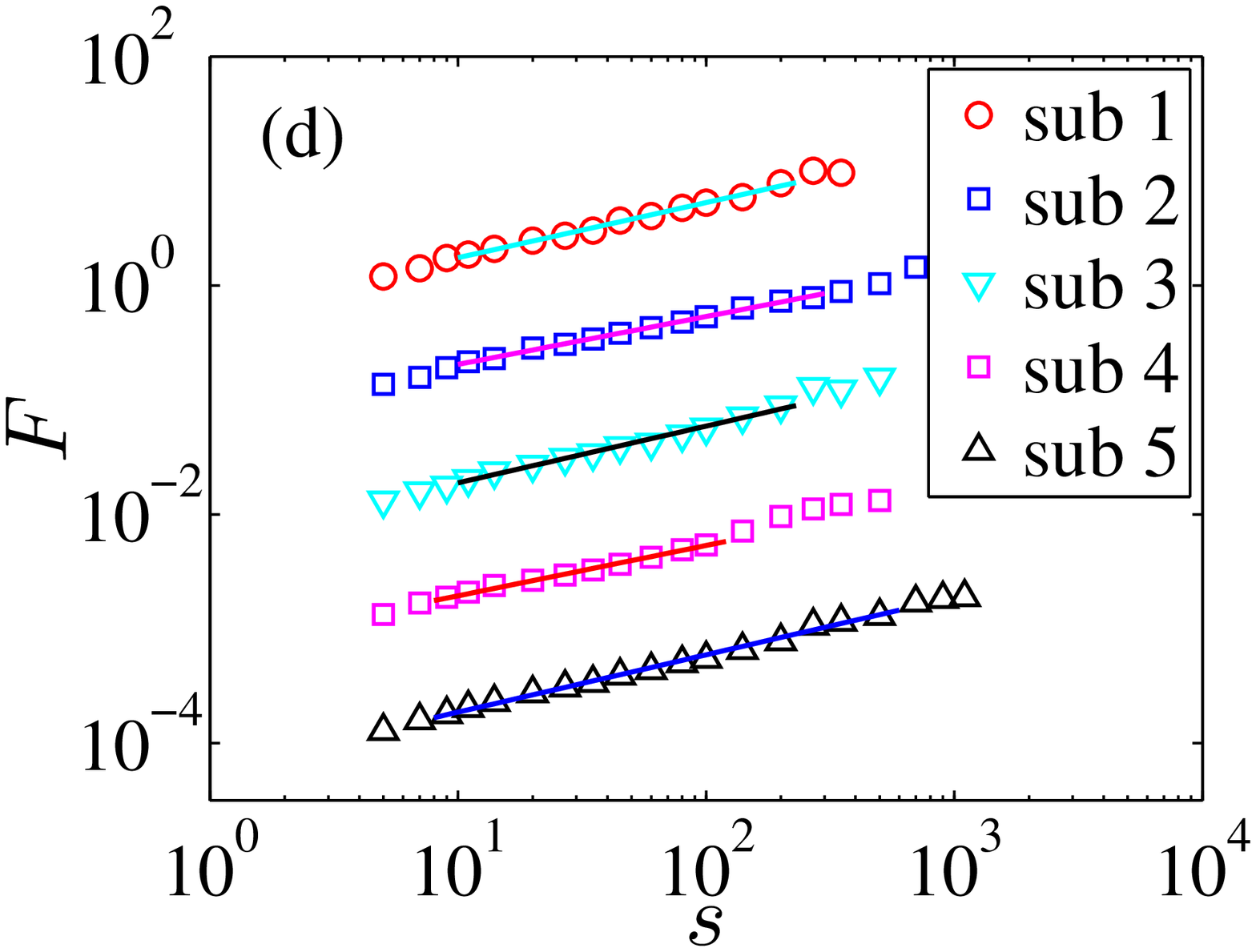}
  \caption{(Color online.) Plots of the fluctuation function $F$ with respect to the box size $s$ for the five sub-series. The curves have been translated vertically for better visibility. (a) DMA with $\theta = 0$. (b) DMA with $\theta = 0.5$. (c) DMA with $\theta = 1$ (d) DFA.}
  \label{Fig:Fut:SubSamples:Fs}
\end{figure}

Each sub-series is injected into the same DMA and DFA analysis procedure as we have done for the whole series. Figure \ref{Fig:Fut:SubSamples:Fs} plots the fluctuation function $F$ as a function of the box size $s$ as open markers for the five sub-series. Again, one can see that there are very nice power-law behaviors, which covers more than one order of magnitude, for all the five sub-series in each panel. The DMA and DFA exponents are estimated by the linear fits to $\ln F$ with respect to $\ln s$ in the scaling ranges, illustrated as solid lines through the open markers. The estimated exponents of the five sub-series are listed in Table~\ref{Tb:Results:SubSamples}.

The proposed statistical test is also performed to verify whether the original sub-series and the shuffled sub-series share the same memory behaviors. We repeatedly reshuffle the original sub-series and calculate the DMA and DFA exponent of the shuffled sub-series till we accumulate an ensemble of 10,000 exponents $H^s$ for each sub-series and each approach. Then the corresponding mean value of shuffled exponents $\langle H^s \rangle$ and $p$-values can be determined, as reported in Table~\ref{Tb:Results:SubSamples}. For comparison, we also list the estimated exponent $H$, the average shuffled exponent $\langle H^s \rangle$, and the $p$-value of the whole series for each approach in the column ``Whole'' of Table~\ref{Tb:Results:SubSamples}.

At the significance level of 0.01, all the four approaches provide the consistent consequences, that the null hypothesis cannot be rejected, for sub-series 1, sub-series 3, and sub-series 5. This confirms that the crude oil market is efficient during those three sub-periods. We also find that the statistical results of DMA with $\theta = 0$ and $\theta = 1$ brightly contradict with the results of DMA with $\theta = 0.5$ and DFA for sub-series 2 and sub-series 4. The DMA with $\theta = 0$ and $\theta = 1$ approaches offers evidence in favor of no long-term correlations in sub-series 2 and sub-series 4, while the DMA with $\theta = 0.5$ and the DFA favor the presence of long-term correlations. From our intuition, this contradiction reveals the presence of certain memory behaviors in sub-series 2 and sub-series 4. One more thing to note is that both sub-periods include the important event of the Gulf war, which implies that the Gulf War had reduced the effectiveness of the crude oil market to certain degree.

\subsection{Moving windows}

To understand the memory behaviors in local samples, we further perform the DMA and DFA analyses on the local series within moving windows. The size of moving windows is defined as 500 and 1000. In each moving window, the original series and the 1000 shuffled series are analyzed by the DMA and DFA algorithm to compute the exponents $H$ and $H^s$, which allows us to proceed our statistical tests. 

When trying to estimated the DMA and DFA exponents through linear regressions of $\ln F$ against $\ln s$ in moving windows, we find that the most tough problem is how to determine the scaling range in each plot of the fluctuation function $F$ versus the box size $s$. As we know, the best way to locate the scaling range is using our naked eyes. However, there are thousands of figures, in which the scaling range need to be determined. We thereby propose an objective strategy to search the scaling range between $F$ and $s$ for each moving window. We require that each scaling range for fitting must contain at least 15 data points, which is close to one order of magnitude, the minimum requirement of the scaling range \citep{Malcai-Lidar-Biham-Avnir-1997-PRE,Avnir-Biham-Lidar-Malcai-1998-Science}. For convenience, we fix the fitting window with a size of 15 data points and slide the fitting window from the first data point till the last data point. The data points in each fitting window are regressed by the linear least-squares estimation and the fitting window with the smallest fitting residual is regarded as the scaling range, whose slope is recorded as the DMA or DFA exponent. For the shuffled series in each moving window, we used the same scaling range as their original series so that they are comparable to each other.

\begin{figure*}[htb]
  \centering
  \includegraphics[width=4.5cm]{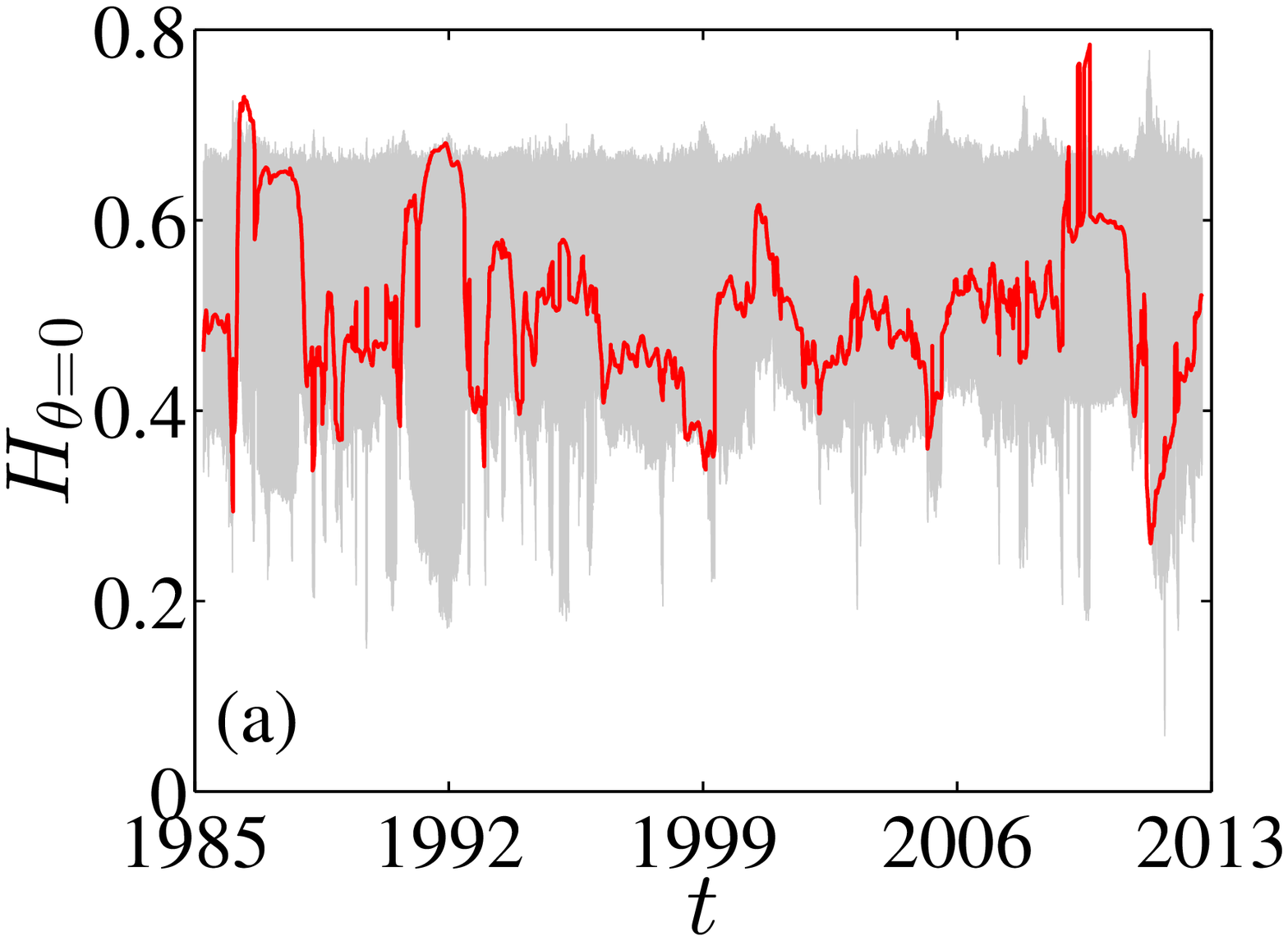}
  \includegraphics[width=4.5cm]{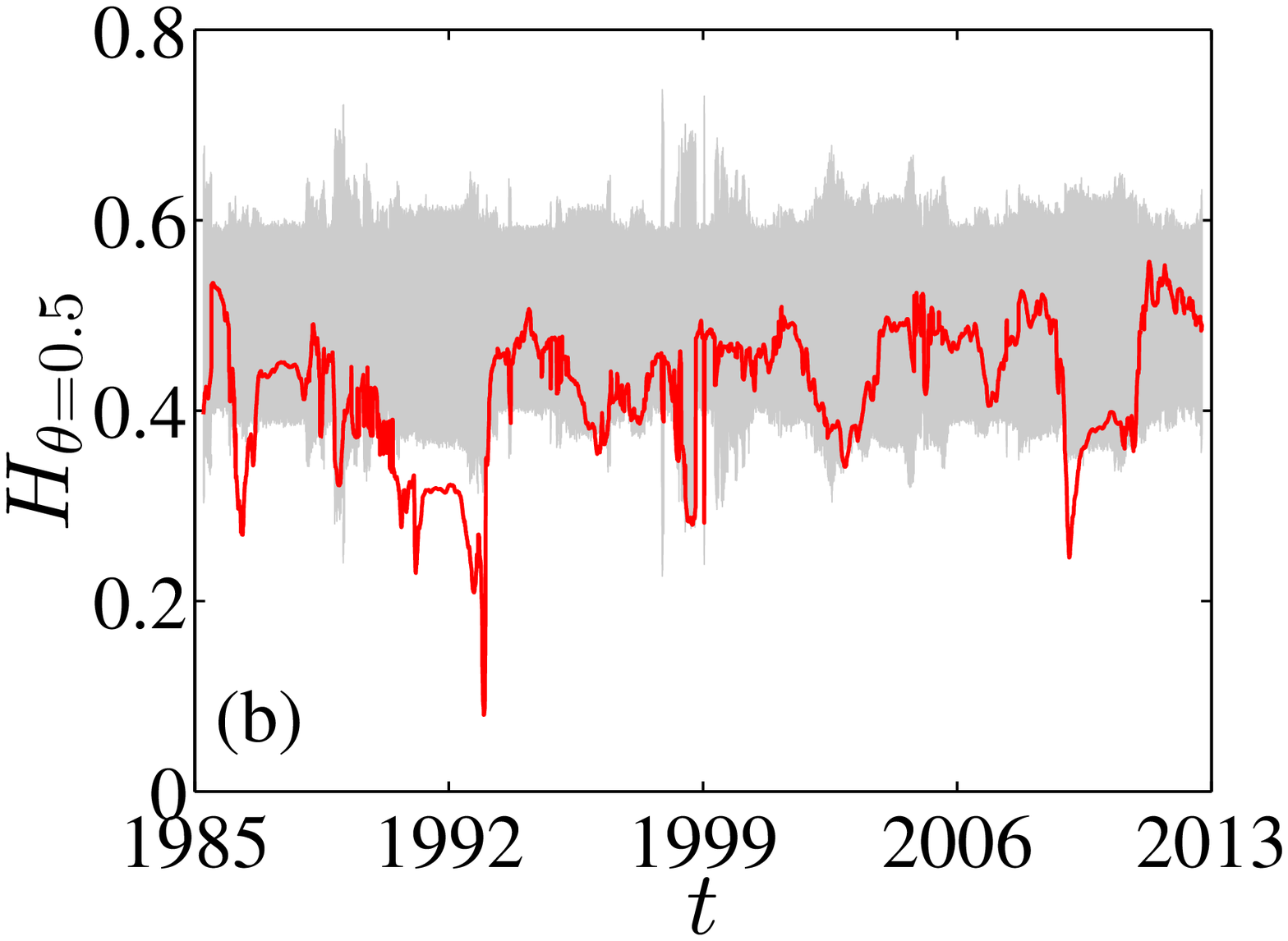}
  \includegraphics[width=4.5cm]{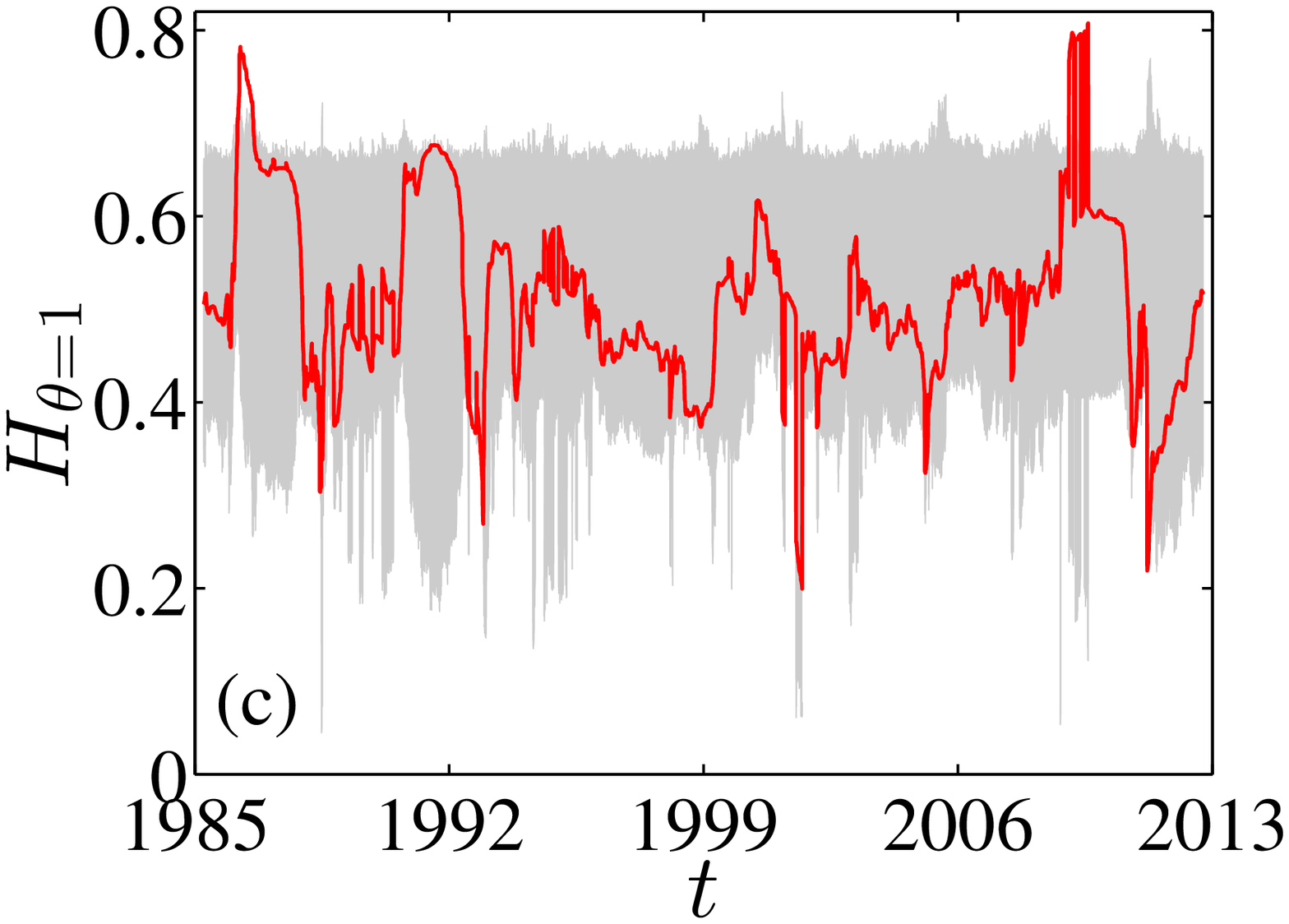}
  \includegraphics[width=4.5cm]{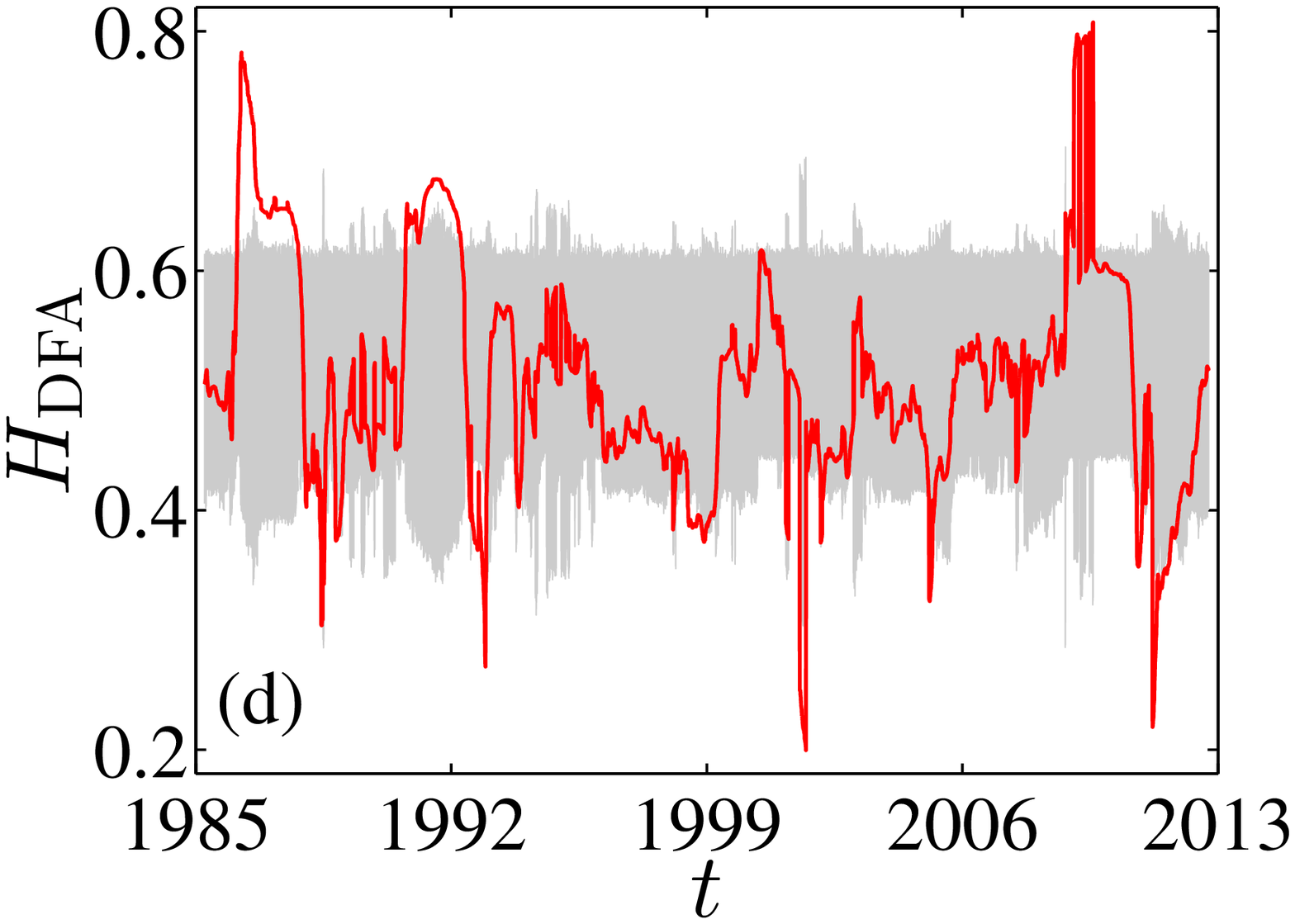}
  \includegraphics[width=4.5cm]{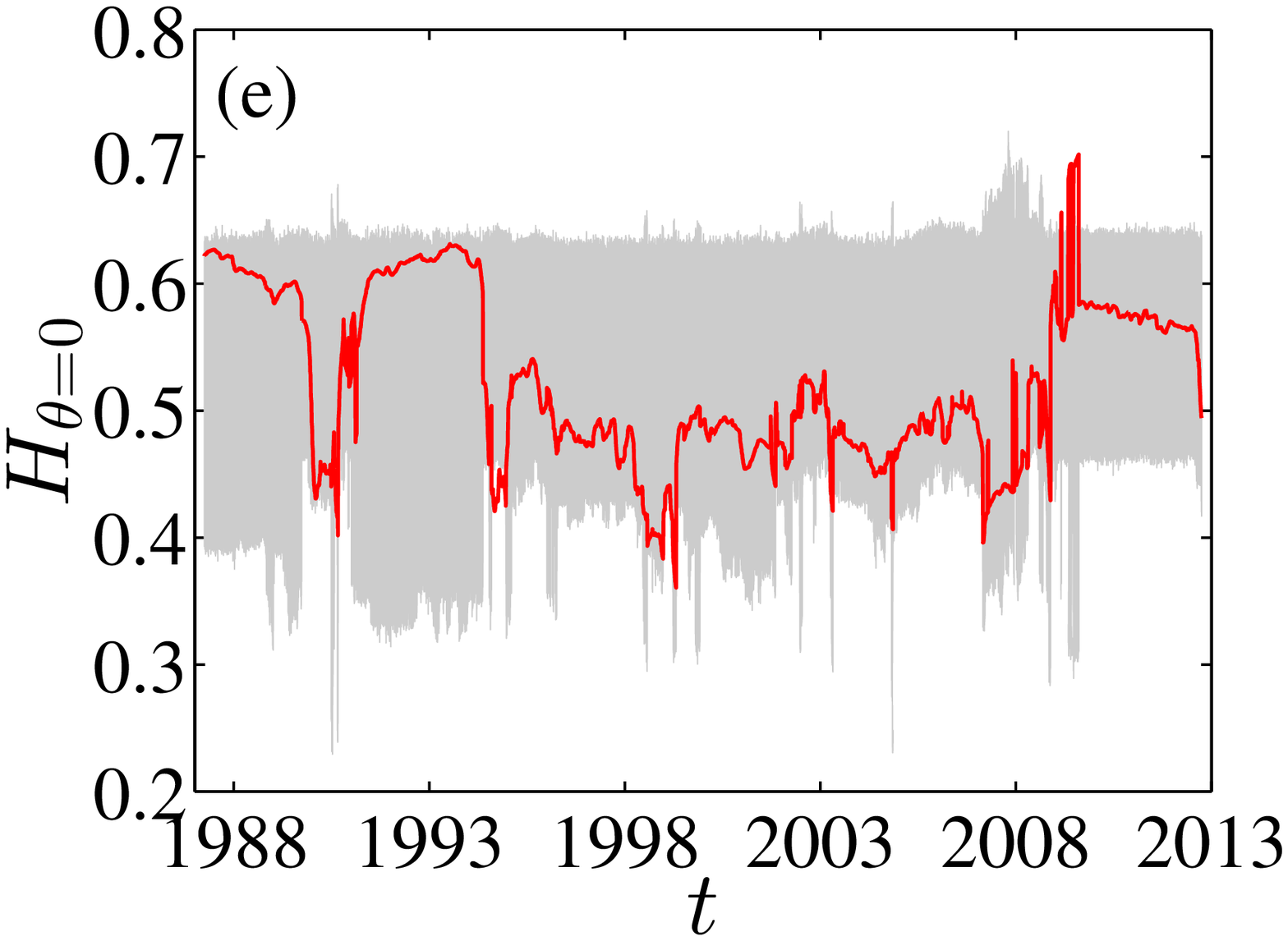}
  \includegraphics[width=4.5cm]{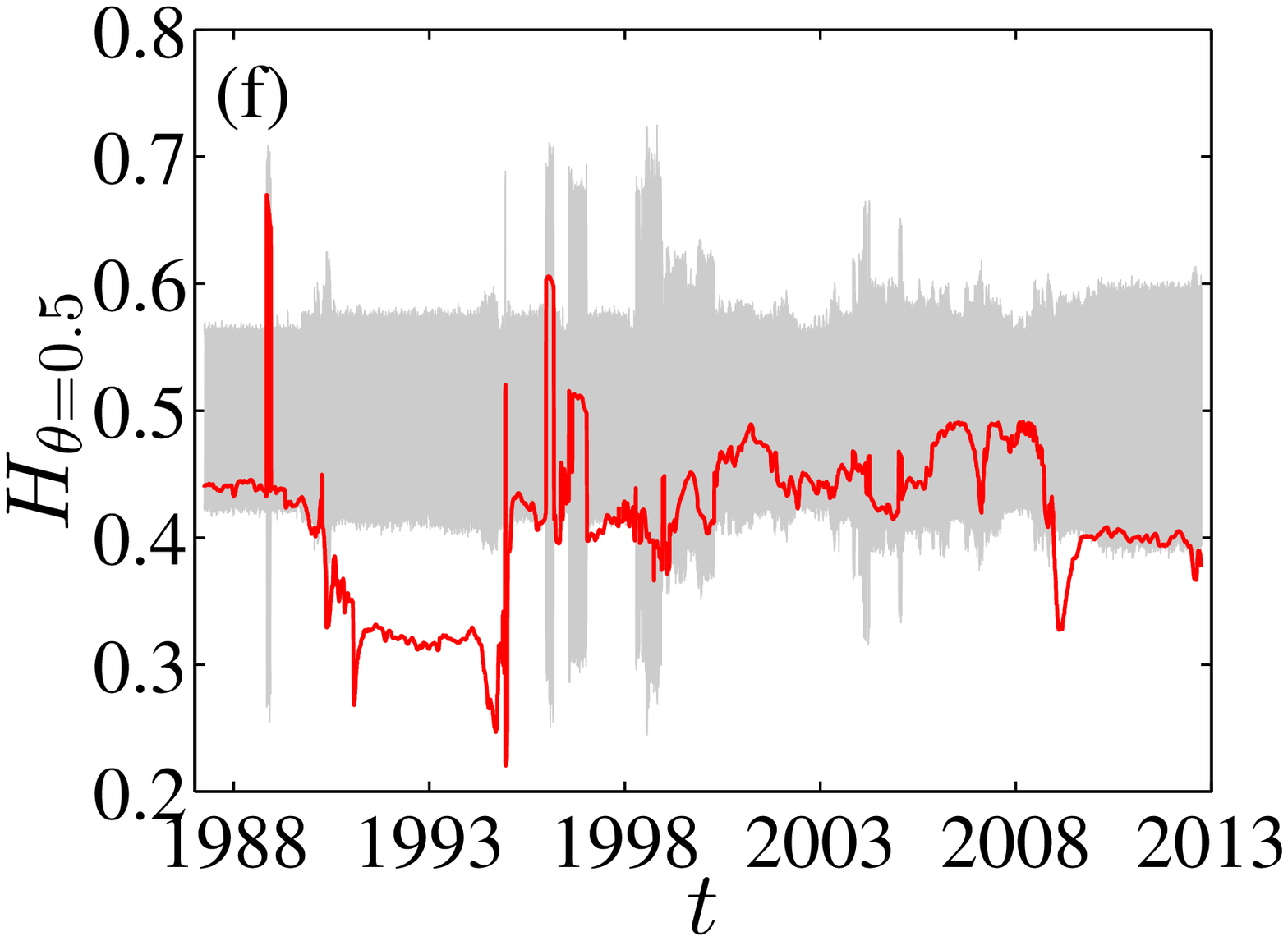}
  \includegraphics[width=4.5cm]{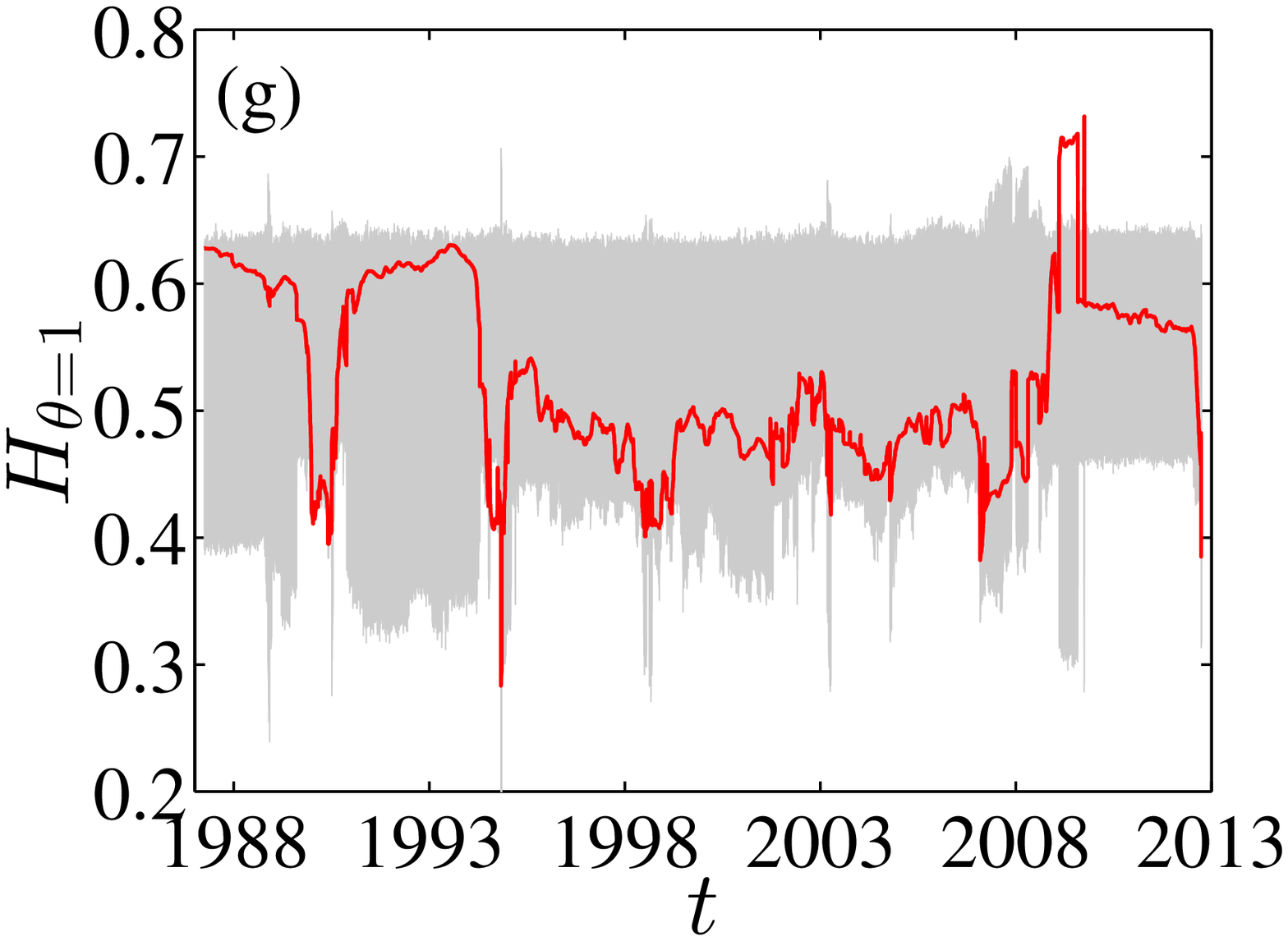}
  \includegraphics[width=4.5cm]{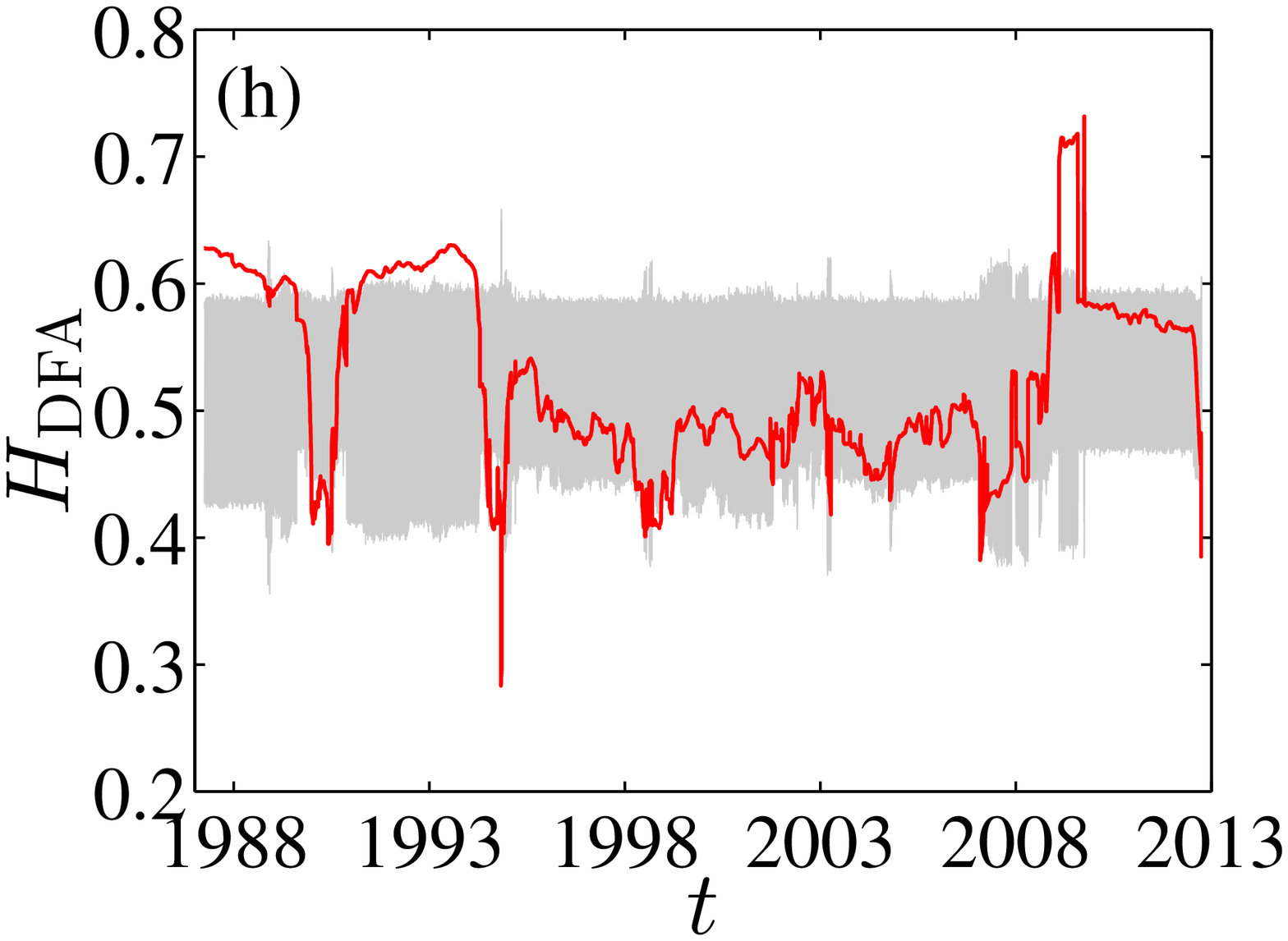}
  \caption{(Color online.) The evolution of estimated Hurst indexes $H$ in moving windows. The shadow areas indicate 2.5/97.5\% quantile range of the Hurst indexes estimated from 1000 shuffled series. Upper panels (a-d) and low panels (e-h) correspond to window size 500 and window size 1000, respectively.}
  \label{Fig:Fut:MovingWindows:H}
\end{figure*}

Figure~\ref{Fig:Fut:MovingWindows:H} illustrates the evolution of the DMA and DFA exponents in moving windows with a size of 500 (upper panels) and 1000 (lower panels) for the four approaches. The date in the abscissa represents the last day of the observations in each moving window. The shadow area corresponds to the 2.5\% and 97.5\% quantiles of the DMA and DFA exponents $H^s$ estimated from the 1000 shuffled series. Although the curve of the exponents $H$ exhibits a strongly fluctuating behavior in each penal, most of the fluctuations are within the shadow area, which indicates that the crude oil market is efficient in most time periods.

We observe that the evolutionary trajectories for BDMA, FDMA, and DFA share remarkable similarity. We further observe that there are some spikes, which exceed the 97.5\% quantile, in plots (a), (c), (d), (e), (g), and (h). In our opinion, these spikes have connections with some turbulent periods, which are related to the oil price crash in 1985, the Gulf War, and the oil price crash in 2008. The phenomena further demonstrate that the effectiveness of the crude oil market in local periods is reduced when the market is in a turbulent state. In these plots, we also find very low $H$ values. Most of them are within the 2.5/97.5\% quantile intervals with a few exceptions in which the scaling ranges are problematic.

\begin{figure}[htb]
  \centering
  \includegraphics[width=4.3cm]{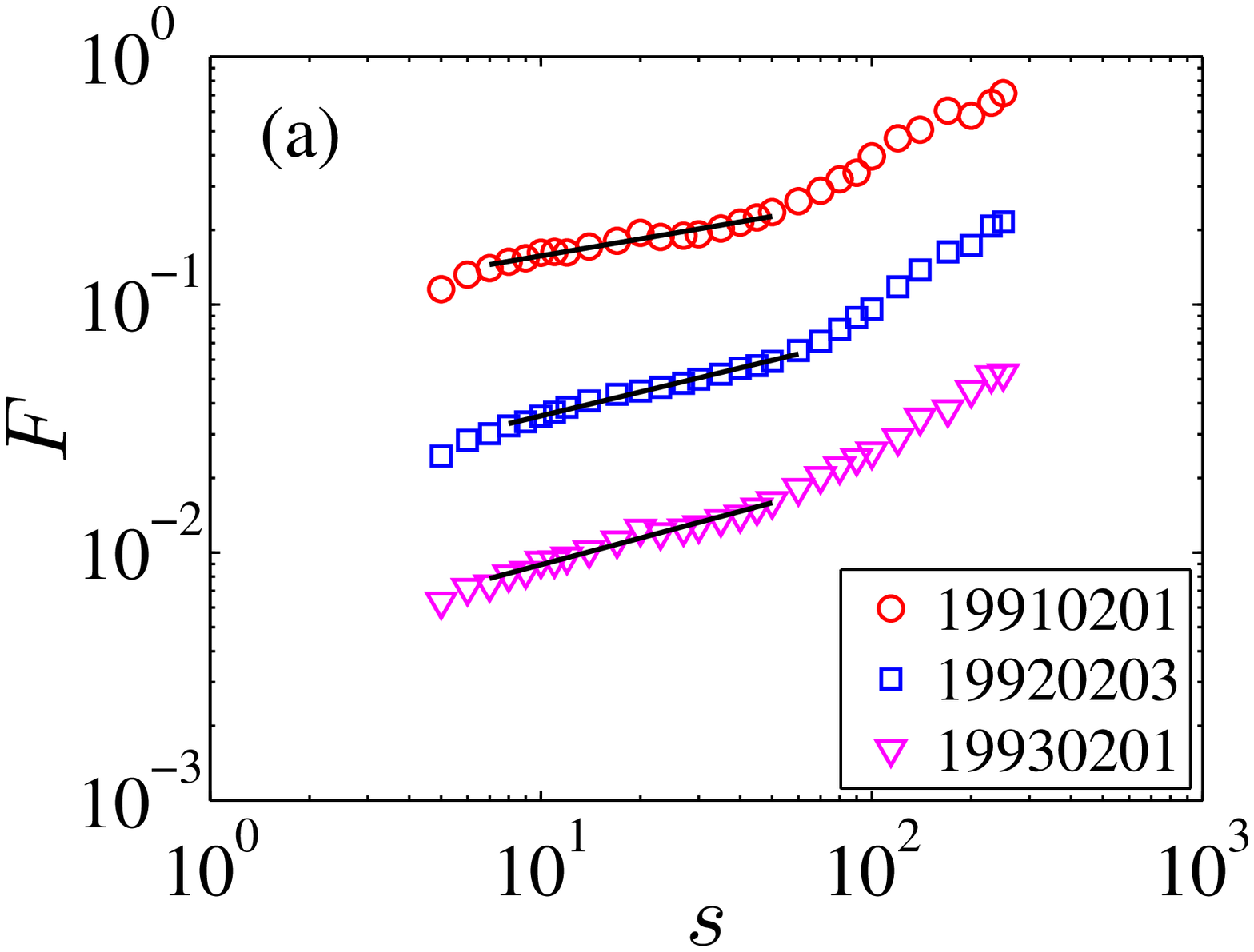}
  \includegraphics[width=4.3cm]{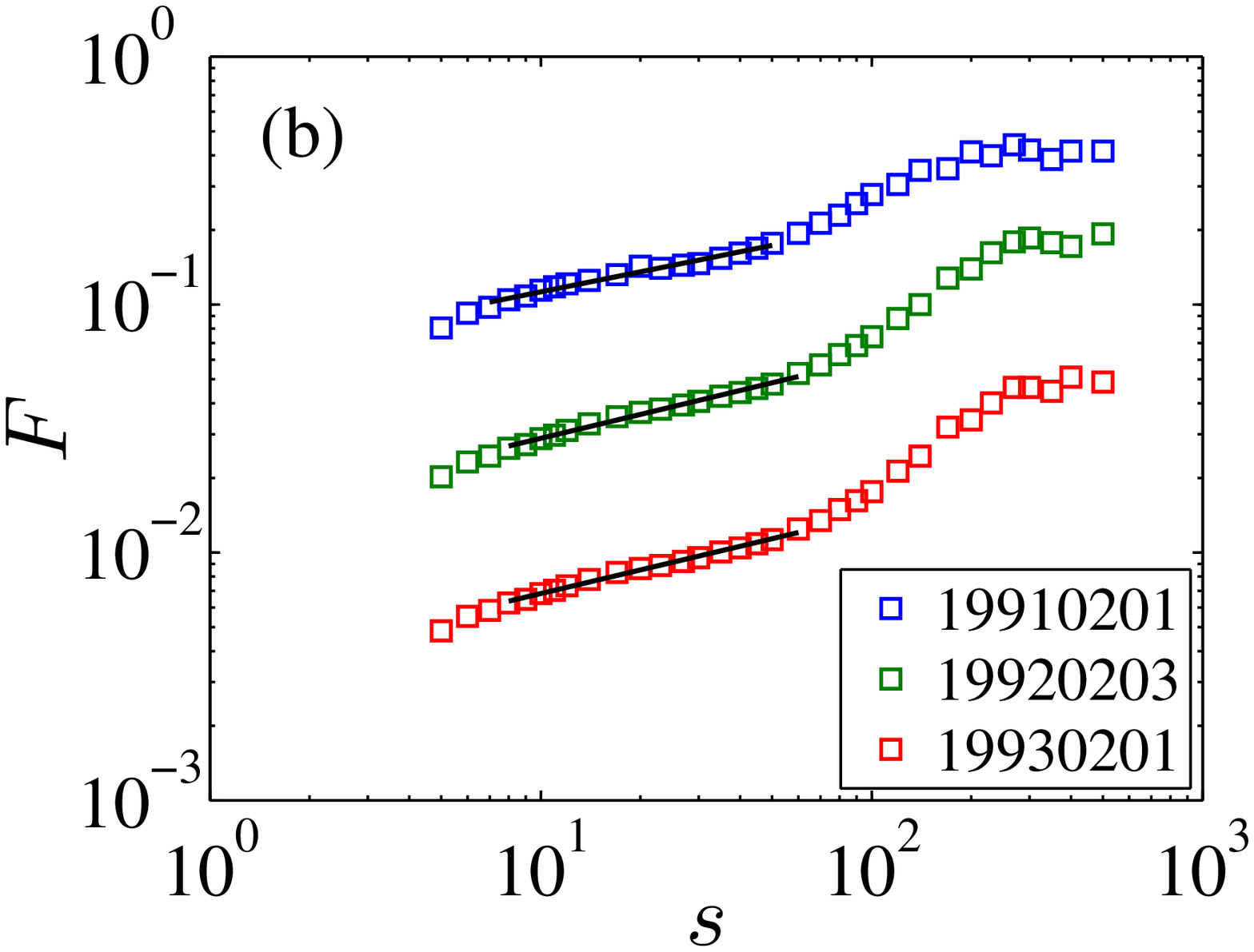}
  \caption{(Color online.) Plots of the fluctuation function $F$ with respect to the box size $s$ for three windows with their ending date on February 1, 1991, February 3, 1992 and February 1, 1993. (a) Window size 500. (b) Window size 1000.}
  \label{Fig:Fut:MovingWindows:Fs:DMA0p5}
\end{figure}

We also notice that the results of DMA with $\theta = 0.5$ exhibit very different behaviors in the period from 1990 to 1993 in plots (b) and (f), when compared with the results of the other three approaches. To find out why this phenomenon occurs, we illustrates in Fig.~\ref{Fig:Fut:MovingWindows:Fs:DMA0p5} the fluctuation function $F(s)$ as a function of the box size $s$ for three typical windows with their ending dates on February 1, 1991, February 3, 1992 and February 1, 1993. Obviously, we can find that there are crossover behaviors for all the three curves in both panels. Our strategic can only locate the scaling range, shown as solid lines, in the range of $s < 50$, which leads to the underestimation of the DMA exponent. However, the crossover phenomena are not observed for the other three approaches in the same window. If we abandon the proposed criterion for determining the scaling ranges and fit the data with $s>50$, the estimates of Hurst indexes become normal and comparable to other approaches. We note that the shape downward spikes in the plots for BDMA, FDMA and DFA have the same behaviors. In conclusion, we argue that these very small values of $H$ do not provide conclusive evidence for the presence of local anti-persistence.

\section{Conclusion}
\label{S1:Conclusion}

In this paper, we performed detailed analyses on the weak-form efficiency of the WTI crude oil futures market by means of DMA and DFA approaches. By using the DMA and DFA exponents as the statistic, we proposed a strict statistical test in the spirit of bootstrapping to check whether the original series exhibit the same memory behaviors as the shuffled series. 

By performing the statistical tests on the whole series, we demonstrated the efficiency of the crude oil market in the investigated period 1983-2012. Our statistical tests have also been carried out on five sub-series in two ways. The two sub-series covering the Gulf War were found to exhibit long-memory behaviors, which means that the Gulf War had an impact of reducing the efficiency of crude oil market. And we also discovered that there were no long-term correlations in other three sub-series. 

By adopting the method of moving windows inspired by \cite{Cajueiro-Tabak-2004a-PA}, we further conducted statistical tests on the observations in each window. This offers us a new insight on the understanding of the evolution of DMA and DFA exponents. We found that there are a lot of fluctuations in the exponent curve; however, most of the fluctuations are inside the 2.5\%-97.5\% quantile  intervals of the exponents of shuffled series. This implies that the crude oil market is efficient in most time periods. The excess of the exponents over the 97.5\% quantile of the shuffled series for the DMA and DFA curves can be connected to some turbulent events, including the oil price crash in 1985, the Gulf war, and the oil price crash in 2008. Our results further revealed that the crude oil market requires a certain time to digest the information. If the investigation time scale is longer than the digesting time, the market looks efficient, and if the investigation time scale is shorter than the digesting time, the market looks inefficient.

\section*{Acknowledgements}

This work was partly supported by National Natural Science Foundation of China (11075054), Shanghai ``Chen Guang'' project (2010CG32), Shanghai Rising Star (Follow-up) Program (11QH1400800), and the Fundamental Research Funds for the Central Universities.


\bibliographystyle{elsarticle-harv}
\bibliography{/home/zqjiang/research/Papers/Auxiliary/Bibliography_FullJournal}







\end{document}